\documentclass[reprint,english,aps,pre,floatfix,showpacs, superscriptaddress]{revtex4-1}

\usepackage{amsmath}
\usepackage{graphicx}
\usepackage{babel}
\usepackage{color}
\usepackage{bm}
\usepackage{subfigure}
\usepackage{dsfont}

\renewcommand{\vec}[1]{\mathbf{#1} }

\begin{document}

\title{Minimal Model for Dynamic Bonding in Colloidal Transient Networks}

\author{Philip Krinninger}
\affiliation{Theoretische Physik II, Physikalisches Institut, Universit\"at Bayreuth, Universit\"atsstra{\ss}e 30, D-95447 Bayreuth,
Germany}
\author{Andrea Fortini}
\affiliation{Theoretische Physik II, Physikalisches Institut, Universit\"at Bayreuth, Universit\"atsstra{\ss}e 30, D-95447 Bayreuth,
Germany}
\affiliation{Department of Physics, University of Surrey, Guildford GU2 7XH, United Kingdom}
\author{Matthias Schmidt}
\email{matthias.schmidt@uni-bayreuth.de}
\affiliation{Theoretische Physik II, Physikalisches Institut, Universit\"at Bayreuth, Universit\"atsstra{\ss}e 30, D-95447 Bayreuth,
Germany}

\date{November 27, 2015, revised: March 10, 2016}


\begin{abstract}
We investigate a model for colloidal network formation using Brownian Dynamics computer simulations. Hysteretic springs establish transient bonds between particles with repulsive core. If a bonded pair is separated by a cutoff distance, the spring vanishes and reappears only if the two particles contact each other. We present results for the the bond lifetime distribution and investigate the properties of the van Hove dynamical two-body correlation function. The model displays crossover from fluid-like dynamics, via transient network formation, to arrested quasi-static network behavior.
\end{abstract}

\maketitle

\section{Introduction}

Network structures are ubiquitous in nature. They influence the properties of many soft matter systems, such as  gels~\cite{REF0}, suspensions~\cite{Puertes, Nature} or entangled polymers~\cite{Green}. At larger length scales, spatial ~\cite{Blair} and force networks~\cite{Snoeijer, Utter} occur in granular matter. 
In living systems, neuronal circuits can be regarded as networks;  the neurones can be identified as nodes, and the synapses serve as links~\cite{Zucker}. 

Many of these examples constitute networks with a static structure, i.e., the position of the nodes, which form the backbone of the network, is fixed in space. Only few of the links between nodes break or form over time. However, there are also {\it transient} networks, where the position of the nodes changes in time. Hence, the general shape of the network changes.
In polymer science the concept of transient networks is well-known~\cite{REF1,REF2,REF3,REF4,REF5} and being used to explain e.g.\ the presence of the rubber plateau in rheological experiments~\cite{deGennes}. Theoretical approaches for transient networks have been developed by {\it Tanaka et al}~\cite{REF1}. In their work, the sticky end-groups of monodisperse polymers form the links.

Transient networks in colloidal systems~\cite{REF6,REF7,REF8}  have been studied in experiments and by numerical simulation. For example colloidal membranes in a magnetic field show effects such as the growth of short chains, cross linking and network formation, induced by many-body polarization interactions between the particles~\cite{Osterman}. A very recent study was aimed at the dynamics of the transient colloidal network itself~\cite{Maier}. In this work, the authors show the influence of the mesh size of the network in the initial state on the mesh dynamics and give an explanation of the shrinking and growing process of the meshes based on the competition of first-order longßrange collective dipolar interactions and short-range second-order dipolar pair correlations.

Dipolar colloidal systems are one of the primary realizations of transient networks. In recent years, progress in the theoretical description of dipolar colloidal gels has been made, supported by extensive molecular dynamics computer simulations~\cite{Referee1, Referee2, Referee3}. These simulation studies on colloidal dumbbells show the crossover from a transient percolated network to a dynamical arrested state as a result of cooling, caused by the rapid increase of bond lifetime of the bonds between different dumbbells at low temperature.

Simulation studies of the influence of solid content on the structure of forming networks of colloidal particles, e.g. the fractal dimension and the bond angle distribution, have been performed~\cite{Hutter}. Patchy colloids~\cite{REF9,REF10,REF11} posses bonding sites on their surface that develop strong short-ranged attractive interactions~\cite{dani_patchy}. The dependence of the network growth on the opening angle of the patches of three-patched colloids has been investigated by {\it Dias et al} very recently~\cite{Dias}. They found different regimes of network formation leading to networks with different structures and sizes. A systematic study of the transition from a fluid to a network in binary mixtures of patchy colloids with varying functionality~\cite{dani_network} has shown the importance of network formation processes for the understanding of transient networks. Transient networks are an intermediate state between a fluid suspension and a fully developed, static, percolated network.

In this article, we present a minimal model for transient network formation in colloidal systems. The model is based on a hysteretic process that describes the formation and annihilation of bonds between colloidal particles with repulsive cores. The bonds form the links of the network, while the particles represent the nodes. The bonds are treated as (linear) springs, inspired by the well-established bead and spring model of polymer physics~\cite{Strobl}. Additionally, the bonding of a pair of particles is based on a hysteretic mechanism: the spring is formed when the surfaces of the two particles touch, and vanishes when the two particles separate above a critical distance, $r_c$. A similar model was proposed for wet granular particles, i.e.\ the minimal capillary model~\cite{Krinninger, Herminghaus}. For wet granular matter dissipative dynamics are considered in molecular dynamics simulations for the collisions, and the interaction between the grains due to capillary bridges is modeled by a constant force. We perform Brownian Dynamics (BD) computer simulations of the minimal model in order to study the deviation of static and dynamic properties from those of a simple suspension of repulsive particles. Moreover, we investigate the network formation properties of the model, from a fluid to a transient network and from a transient to a static network.  

The paper is organized as follows. In Sec.~\ref{chap:model-method} we introduce the model and the simulation technique, as well as the van Hove dynamic correlation function which we use as a mean to characterize the system. In Sec.~\ref{chap:results} we present our results. First, we study statistical properties of the bonding in Sec.~\ref{sec:bonds}. In particular, we are interested in the lifetime of bonds from formation to annihilation and the corresponding probability distribution. We then focus on static properties, namely the percolation transition and the fractal dimension of percolating clusters of colloids in Sec.~\ref{sec:perc}. In Sec.~\ref{sec:hove}, we give an overview of the detailed studies of the van Hove function as a function of density $\rho$, the bond strength $k$, and the correlation time. We investigate the change of correlation with increasing the bond strength, up to the point where the system is no longer fluid. This crossover manifests itself in a non-Gaussian shape of the self part of the van Hove function and is discussed in detail in Sec.~\ref{sec:ngauss}. In Sec.~\ref{sec:conclusion} we conclude and give an outlook to possible future work within the framework of the proposed model.

\section{Model and Method}
\label{chap:model-method}
We consider a three-dimensional system of $N$ interacting, spherical Brownian 
particles with spatial coordinates 
$\vec r_i$, $i= 1 \dotsc N$.
We neglect hydrodynamic interactions and describe the dynamics with the 
overdamped Langevin equation
\begin{align}\label{full_langevin}
\dot{\vec r}_i =  \gamma^{-1}\vec F_i + \boldsymbol \xi_i(t),
\end{align}
where $\gamma$ is the friction coefficient. The deterministic force on particle $i$ is generated from the 
total potential energy $U_N$ according to $\vec F_i\!=\!-\nabla_i U_N$, where $\nabla_i$ denotes the derivative with respect to $\vec r_i$.
The stochastic random force $\gamma \boldsymbol \xi_i(t)$ is Gaussian distributed with zero mean and autocorrelation	
$\langle \boldsymbol \xi_i(t) \boldsymbol \xi_j(t')\rangle=2D_0\mathds{1}\delta_{ij}\delta(t-t')$, where 
$D_0$ is the Stokes-Einstein diffusion coefficient, $\mathds{1}$ denotes the $3\times 3$ unit matrix, $\delta_{ij}$ is the Kronecker delta and $\delta(\cdot)$ indicates the Dirac distribution.

\begin{figure}
\centering
\includegraphics[width=8cm]{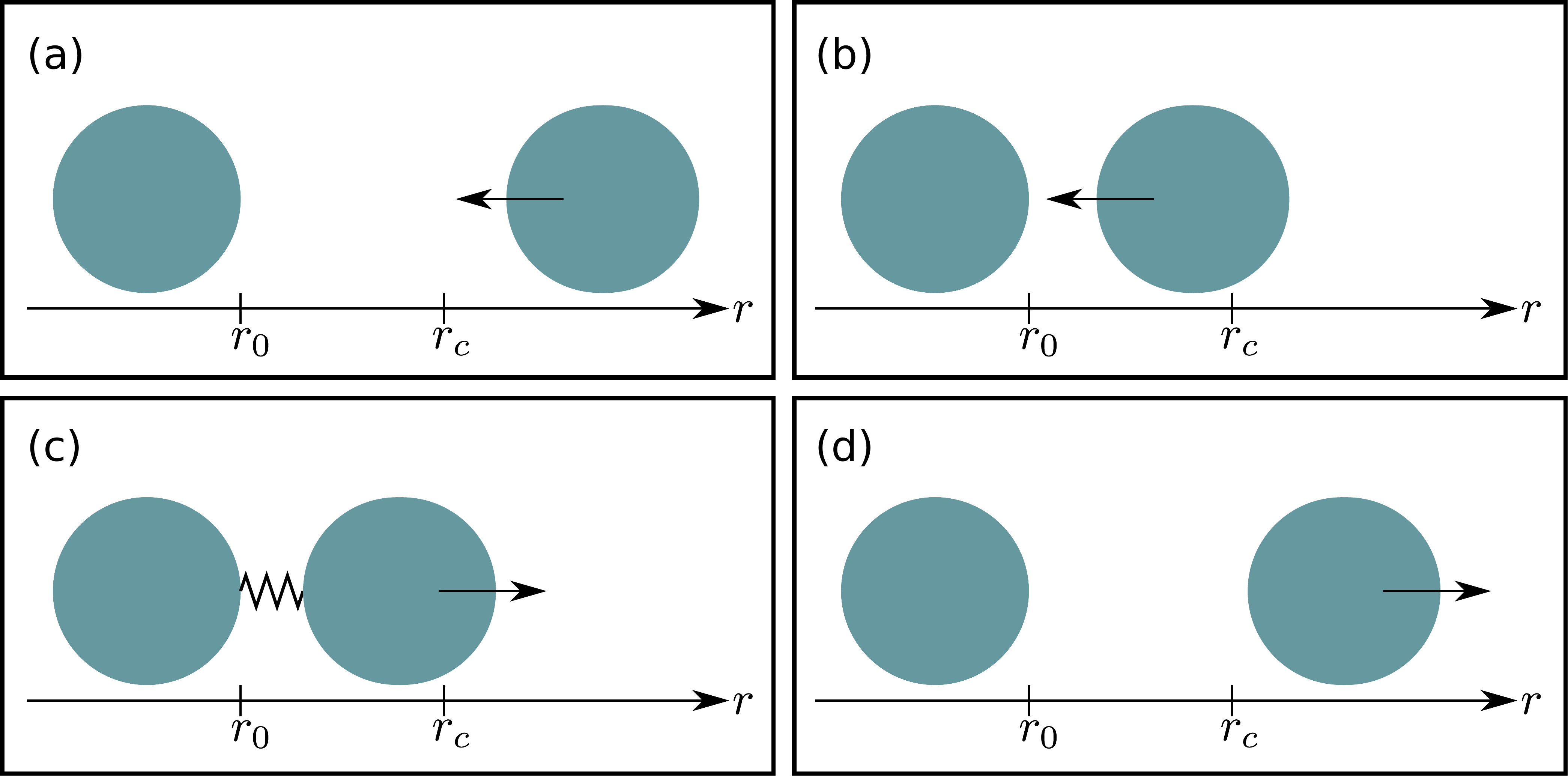}
\caption{Sketch of the forming and vanishing of a bond between two particles. The equilibrium distance of the spring, $r_0$, is the contact distance of the particles. The arrows indicate the direction of the motion of the particle. (a) No interaction because the particles are too far apart from each other. (b) The distance of the particles is smaller than $r_c$, but still no interaction because no previous contact between the particles has occurred. (c) After the contact, the bond is formed and remains as long as the distance between the particles is smaller than $r_c$. (d) The spring vanishes because the particles are too far apart from each other.\label{fig:model}}
\end{figure}

The interaction potential $U_N$ is a pairwise, particle-particle interaction potential tailor-made for network formation. It combines a repulsive interaction $U_{\rm REP}$ with a harmonic potential $U_{\rm S}$ for the links between the particles: \begin{align}
U_N= \frac{1}{2}\sum_{i=1}^N \sum_{j=1 \atop j\neq i}^N (U_{\rm REP}(r_{ij})+\nu_{ij} U_{\rm S}(r_{ij})),
\end{align}
where $r_{ij}=|\vec r_i - \vec r_j|$ and $\nu_{ij}=0,1$ is a bonding degree of freedom that determines whether particles $i$ and $j$ interact at time $t$ via a spring ($\nu_{ij}=1$) or not ($\nu_{ij}=0$). The linking, and hence the value of $\nu_{ij}$, is history dependent, illustrated by Fig.~\ref{fig:model}: When the surfaces of two particles $i$ and $j$ touch, they become bonded by a spring ($\nu_{ij}=1$). When the particles separate above a critical distance $r_c$ the bond vanishes ($\nu_{ij}=0$). For the repulsive core we use $U_{\rm REP}=\epsilon( \sigma / r_{ij})^{12}$, where $\epsilon$ is the unit of energy, $\sigma$ is the particle diameter. $U_{\rm REP}$ is cut off and shifted at $r_{\rm cut} /\sigma =1.01$ to avoid discontinuities in the interaction potential. The harmonic potential is $U_{\rm S}(r_{ij})=\frac{k}{2}(r_{ij}-\sigma)^2$, where $k$ is the stiffness of the spring determining the bond strength. Here the equilibrium distance of the spring is chosen to be the core size of the repulsive interaction, $\sigma$. 

We carry out Brownian dynamics (BD) simulations with a fixed time step of $\delta t/\tau_B=8\times 10^{-5}$, with the Brownian time $\tau_B=\sigma^2/D_0$. The fundamental units of the system are $\sigma$, $\gamma$ and $\epsilon$. All simulations are performed at a reduced temperature of $k_B T/\epsilon=2$, where $k_B$ is the Boltzmann constant, and a fixed critical distance of the hysteretic spring of $r_c/\sigma=1.5$. The particles are placed in a cubic, periodic box which side length $L=(N/\rho)^{1/3}$, where $\rho=N/V$, with $V$ being the volume of the simulation cube. We investigate the properties of the system as a function of the density $\rho$, and the strength of the hysteretic links, $k$. We carried out simulations with a density of $\rho \sigma^3 = 0.1$ to 0.5 in steps of 0.05 and $\rho \sigma^3=0.6$, and bond strengths of $k\sigma^2/\epsilon=0$, 10, 20, 40, 70. Furthermore, for $k\sigma^2/\epsilon=40$ and 70 the densities $\rho/\sigma^3=0.01$ and 0.05 were considered.

\subsection{van Hove Correlation function}

We characterize the dynamical correlations using the van Hove function $G(\vec r, t)$~\cite{vhove:vhove, hansen:mcdonald}. It  characterizes the spatial and the temporal distribution of pairs of particles, as is relevant for fluid states. $G(\vec r, t)\rm d \vec r$ can be interpreted as the number of particles $j$ in a volume element $\rm d \vec r$ at position $\vec r$ under the condition that there was a particle $i$ at the origin at time $t=0$. $G(\vec r, t)$ is related to the intermediate scattering function $F(k,t)$, which is measurable in x-ray or neutron scattering experiments, via spatial Fourier transform, and to the dynamic structure factor $S(k,\omega)$ via spatial and temporal Fourier transform. Further motivation for considering $G(\vec r,t)$ stems from recent theoretical progress in formulating an exact generalization of the Ornstein-Zernike relation to nonequilibrium situations~\cite{NOZ1,NOZ2}. Here dynamical correlation functions are related to functional derivatives of a generating (free power dissipation) functional~\cite{PFT}. An alternative theoretical description rests on the dynamical test particle limit~\cite{thevanhove,ajarcher}, which was recently treated within the power functional approach~\cite{TPL}.

The van Hove function is defined as~\cite{vhove:vhove, hansen:mcdonald}
\begin{equation}
G(\vec r,t) = \frac{1}{N} \left\langle \sum_{i=1}^N \sum_{j=1}^N \delta (\vec r + \vec r_j(0) - \vec r_i (t)) \right\rangle ~,
\label{eq:vanhove}
\end{equation}
where $\langle \cdot \rangle$ indicates the ensemble average, $\delta ( \cdot )$ is the (three-dimensional) Dirac delta function. It is possible to split $G(\vec r ,t)$ into a self and a distinct part. In the first case the double sum is restricted to $i=j$ and $G_{\rm self}(\vec r, t)$ describes the average motion of a particle that was at the origin at the initial time. The distinct part, $G_{\rm dist}(\vec r, t)$, where $i \neq j$, represents the remaining $N-1$ particles, considering that any arbitrary particle $j$ was located at $\vec r_j=0$ at $t=0$. Therefore
\begin{equation}
\begin{split}
G ( \vec r, t) =& \frac{1}{N} \left\langle \sum_{i=1}^N \delta ( \vec r + \vec r_i(0) - \vec r_j(t) ) \right\rangle \\
 &+ \frac{1}{N} \left\langle \sum_{i,j=1 \atop i\neq j}^N \delta ( \vec r + \vec r_j (0) - \vec r_i(t)) \right\rangle \\
\equiv& G_{\rm self}(\vec r, t) + G_{\rm dist}(\vec r, t)~.
\end{split}
\label{eq:vanhove_self_distinct}
\end{equation}
Hence, the self-part describes the dynamics of only one tagged particles, while $G_{\rm dist}$ represents the remaining $N-1$ particles. Therefore the normalization of the self and distinct parts is 
\begin{align}\label{eq:vanhove_norm1}
\int \text d \vec r~ G_{\rm self}(\vec r,t)&=1~, \\
\int \text d \vec r~ G_{\rm dist}(\vec r, t) &= N-1~.
\label{eq:vanhove_norm2}
\end{align}
The initial time behavior for $t=0$ of $G(\vec r, t)$ is given by 
 \begin{equation}
\begin{split}
G( \vec r, 0) =& \delta (\vec r) 
+ \frac{1}{N} \left\langle \sum_{i,j=1 \atop i\neq j}^N \delta (\vec r + \vec r_j(0) - \vec r_i (t))\right\rangle \\
=& \delta (\vec r) + \rho g(\vec r) ~,
\end{split}
\label{eq:vanhove_t-0}
\end{equation} 
where $g(\vec r)$ is the pair correlation function. Hence $G_{\rm self}(\vec r,0) = \delta (\vec r)$ and $G_{\rm dist}(\vec r,0) = \rho g(\vec r)$. As time passes, the $\delta$-function broadens into a bell-shaped curve, and the peaks of $G_{\rm dist}$ decrease and disappear. For $t\to \infty$ the correlation vanishes and $G(\vec r, t)$ becomes a constant, where $G_{\rm self}(\vec r, t \to \infty) = 0$, and $G_{\rm dist}(\vec r, t\to \infty) = \rho$.

One important property for a homogeneous bulk fluid is that the van Hove function only depends on the distance $r=|\vec r|$, because of the isotropy: $G(r,t) = G_{\rm self}(r,t) + G_{\rm dist}(r, t)$.

The free motion of one single particle in Brownian dynamics is a random walk, hence free diffusion occurs with the diffusion coefficient $D_0$. In this situation the self-part of the van Hove function is given by the solution of the diffusion equation~\cite{hansen:mcdonald, thevanhove}:
\begin{align}
\frac{\partial}{\partial t}G_{\rm self}(r,t) = D_0 \nabla ^2 G_{\rm self}(r,t)~,
\label{eq:diff}
\end{align}
which is is
\begin{align}
G_{\rm self}(r,t) = (4\pi D_0 t)^{-3/2} \exp\left( -\frac{r^2}{4 D_0 t} \right)~.
\label{eq:gaussian}
\end{align}
For the many-body system this expression is exact for $\rho \to 0$, as the interactions between the particles can be neglected in this limit. In systems with finite density Eq.~(\ref{eq:gaussian}) is an approximation where $D_0$ becomes an effective diffusion coefficient, which is a function of density, $D(\rho)$. Increasing the interaction between the particles further, i.e.\ by strong bonding in the current work, can lead to the shape of $G_{\rm self}(r,t)$ deviating from a Gaussian. The deviation can be quantified (in three dimensions) by the non-Gaussian parameter 
\begin{align}
\alpha_2 (t)=\frac{3 \langle r^4(t) \rangle}{5 \langle r^2(t) \rangle ^2}-1~,
\label{eq:ngauss}
\end{align}
where $\langle r ^{\mu}(t) \rangle = \int \text d \vec r r^{\mu} G_{\rm self}(r, t)$ is the $\mu$-th spatial moment of $G_{\rm self}(r,t)$~\cite{ngauss:kob, ngauss:rahman}. For a strict Gaussian $\alpha_2=0$.

\subsection{Mean First Passage Time}
A simple theoretical description of bond lifetime is given by the mean first passage time $\tau$ for a particle in an external potential. In the framework of the Kramer's problem in one dimension it is possible to calculate $\tau$ from the adjoint Smoluchowski equation~\cite{Zwanzig}. In this approach the motion of a single Brownian particle in an external potential is considered. The purpose is to calculate the mean time it takes the particle to escape the potential, i.e.\ when it reaches a certain end point. The starting position of the particle, $x$, is between a reflective barrier, located at the point $a$ and the end point $b$, with $a<x<b$. With these assumptions one can calculate the mean first passage time in one dimension as a function of the starting position $x$~\cite{Zwanzig}.

In order to adopt this theory to our model, we consider a pair of bonded particles in three dimensions. One particle serves as the origin of the coordinate system and the other particle escapes the harmonic potential generated by the bond between the colloids. Therefore  we choose for the external potential $U=U_S$. Furthermore, we generalize the calculation of $\tau$ to three dimensions, starting with the three dimensional adjoint Smoluchowski equation
\begin{align}
D \exp\left( \frac{U(\vec r)}{k_BT} \right) \nabla \cdot \left[ \exp\left( -\frac{U(\vec r)}{k_BT} \right) \nabla \tau(\vec r_0) \right] = -1 ~,
\end{align}
where $\vec r_0$ is the starting point of the particle in the harmonic potential.
Because the total interaction potential only depends on the distance between the particles it can be written as 
\begin{align}
	D \exp\left( \frac{U(r)}{k_BT} \right) \frac{1}{r^2} \frac{\partial}{\partial r} \left[ r^2 \exp\left( -\frac{U(r)}{k_BT} \right) \frac{\partial \tau(r_0)}{\partial r} \right] = -1 ~.
\end{align}
Integrating twice leads to the mean first passage time $\tau$:
\begin{align}
	\tau(r_0) = \frac{1}{D} \int_{r_0}^{r_b} \text d y \frac{1}{y^2} \exp\left( \frac{U(y)}{k_BT} \right) \int_{r_a}^y \text d z z^2 \exp\left( -\frac{U(z)}{k_BT} \right) ~ ,
	\label{eq:meanfirstpassagetime_3d}
\end{align}
with $r_0$ is the starting position, $r_a$ is the position of the reflecting barrier and $r_b$ is the end position. In the current work the values for $r_0$, $r_a$ and $r_b$ are $r_0/\sigma=1$, $r_a/\sigma =1$ and $r_b/\sigma=1.5$, and for $D$ we choose $D=2D_0$, as the origin is given by a diffusively moving particle, see e.g.\ Ref.~\cite{0295-5075-102-2-28011}. Hence, Eq.~(\ref{eq:meanfirstpassagetime_3d}) is only exact if there is one pair of bonded particles, i.e.\ $\rho \to 0$. At finite densities the collisions with surrounding particles lead to a different lifetime of the bonds.

\section{Results}
\label{chap:results}

\subsection{Bond statistics}
\label{sec:bonds}
\begin{figure}
\centering
\includegraphics[width=8cm]{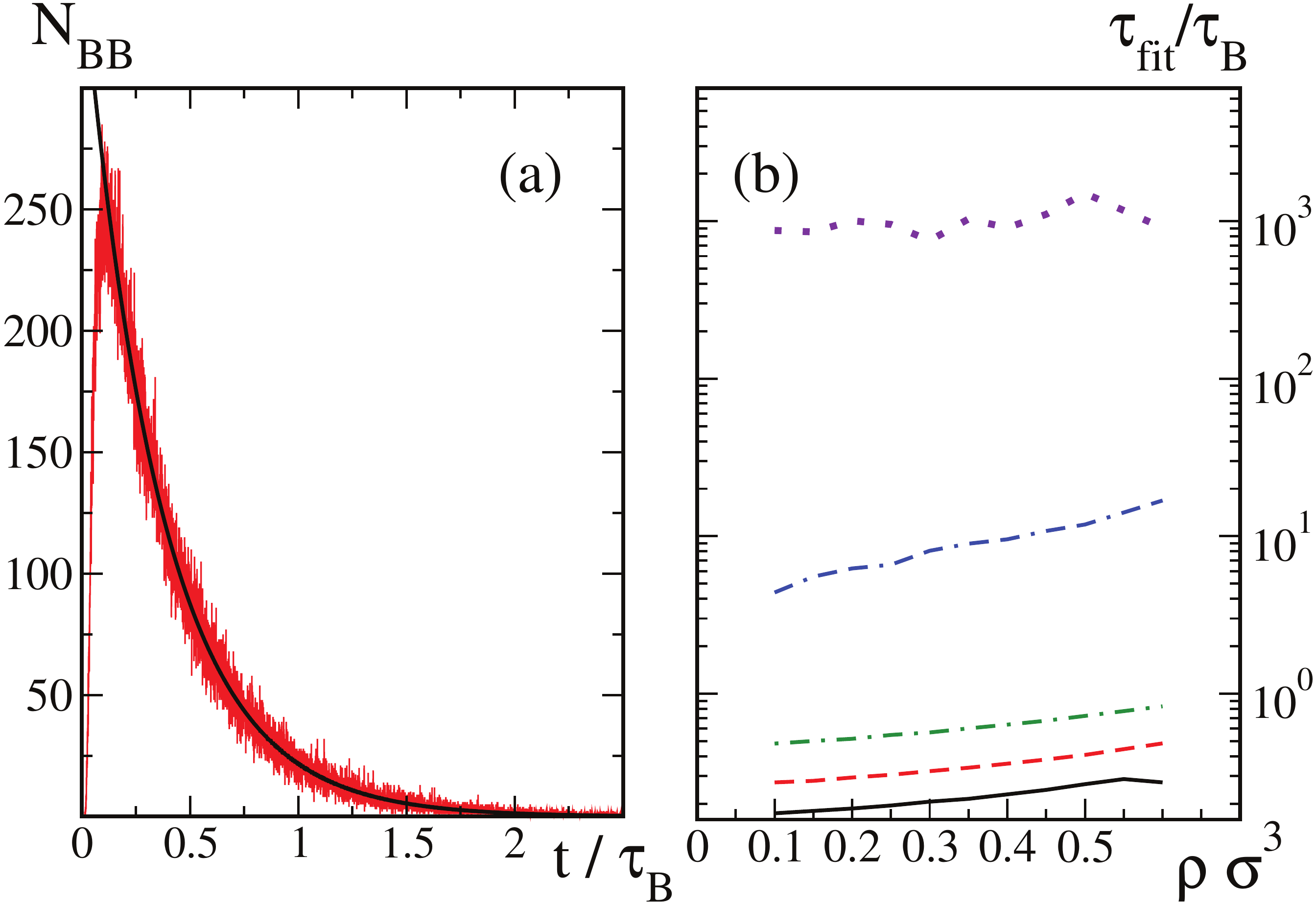}
\caption{Bond lifetime statistics: (a) Histogram of the number of bonds that break over time, $N_{BB}$ for parameters $\rho \sigma^3 = 0.4$ and $k \sigma^2 /\epsilon =10$. The black line is a fit to an exponentially decaying function. (b) Fit parameter $\tau_{\rm life}$ as a function of density, for different bond strengths: $k\sigma^2/\epsilon=0$ (black solid line), 10 (red dashed line), 20 (green dashed-dotted line), 40 (blue dashed-dashed-dotted line) and 70 (purple dotted line).\label{fig:bonds}}
\end{figure}

We start by investigating the properties of the dynamic bond formation process. We consider the time scale on which a spring is active, i.e.\ how much time passes between the formation and the annihilation of a certain bond. We study this process by varying systematically the mean density and the bond strength. In Fig.~\ref{fig:bonds}(a), we present a histogram of the bond lifetime for the parameters $\rho \sigma^3 = 0.4$ and $k \sigma^2 /\epsilon=10$, where $N_{BB}(t)$ marks the number of broken bonds after they existed for a time $t$. The black curve is a fit to the function $N_{BB}(t)=N_0 \exp(-t/\tau_{\rm life})$, where $\tau_{\rm life}$ is the average lifetime of the bond. Results for $\tau_{\rm life}$ for further parameters are shown in Fig.~\ref{fig:bonds}(b). We observe that either increasing $\rho$ or $k$ leads to an increase of the lifetime. A harder spring (increasing $k$) leads to a stronger attraction between the bonded pairs, which makes it harder for the particles to separate from each other above the critical distance $r_c$, resulting in a increased bond lifetime. The increase in density causes an increase in the number of collisions, and therefore makes it more unlikely for a particle to separate from its bonded partner, increasing the lifetime.

The results for the mean first passage time, $\tau$ and $\tau_{\rm life}$, are summarized in Table~\ref{tab:passagetimes}. Comparing these values with the simulation results, we find some  discrepancies, which are entirely expected. First, the calculation Eq.~(\ref{eq:meanfirstpassagetime_3d}) neglects the repulsive core interaction, which is a small error, as the cut-off length is chosen rather short, compared to the maximal possible spring length. Second, Eq.~(\ref{eq:meanfirstpassagetime_3d}) is only exact for $\rho \to 0$. Third, there is statistical error. Especially for $k \sigma^2 /\epsilon=40$ and 70, the particles get very sticky, and bond breaking becomes rare, making the statistical error the most dominant in these systems. For $k \sigma^2 /\epsilon = 0$ and $k \sigma^2 /\epsilon = 10$ the accordance of $\tau_{\rm life}$ at $\rho \sigma^3 = 0.1$ with the calculated mean first passage times is quite good. But as the density increases, the discrepancy between the theoretically predicted values and the one sampled from simulated data increases, as expected. For $k \sigma^2 /\epsilon = 20$ in the low density regime ($\rho \sigma^3 =0.1$) the deviation from the theory is higher than in the cases before. In the case of $k \sigma^2 /\epsilon = 40$ the comparison between calculation and simulation is only reasonable for low densities, where we find $\tau_{\rm life}/\tau_B=3.589$ for $\rho \sigma^3 = 0.01$ and $\tau_{\rm life}/\tau_B=3.786$ for $\rho \sigma^3 = 0.05$. For increasing density the differences between calculation and simulation increase further. As mentioned above, the comparison for $k \sigma^2 /\epsilon = 70$ is hardly possible and the differences between the values is large even in the low density case, where the simulations give $\tau_{\rm life}/\tau_B=822.596$ for $\rho \sigma^3 = 0.01$ and $\tau_{\rm life}/\tau_B=519.481$ for $\rho \sigma^3 = 0.05$. 
Despite quantitative discrepancies with the simulations, the theory captures the correct trend of increasing relaxation times for increasing density and bond strength.

\begin{table} [htbp] \centering 
\caption{Mean first passage times $\tau$ for different bond strengths $k$ as calculated by Eq.~(\ref{eq:meanfirstpassagetime_3d}) and simulation results for $\rho \sigma^3 =0.1$ and 0.6.\label{tab:passagetimes}}
\begin{tabular}{r|ccccc}
$k  \sigma^2/\epsilon$ & 0 & 10 & 20 & 40 & 70 \\ \hline
$\tau/\tau_B$ from Eq.~(\ref{eq:meanfirstpassagetime_3d}) & 0.388 & 0.472 & 0.588 & 1.000 & 2.764 \\ \hline
$\tau_{\rm life} /\tau_B$ for $\rho \sigma^3 =0.1$ & 0.174 & 0.273 & 0.482 & 4.389 & 872.476 \\
$\tau_{\rm life} /\tau_B$ for $\rho \sigma^3 =0.6$ & 0.273 & 0.483 & 0.831 & 16.839 & 908.690 \\
\end{tabular}
\end{table}

\subsection{Percolation and fractal dimension}
\label{sec:perc}
\begin{figure}
\centering
\includegraphics[width=8cm]{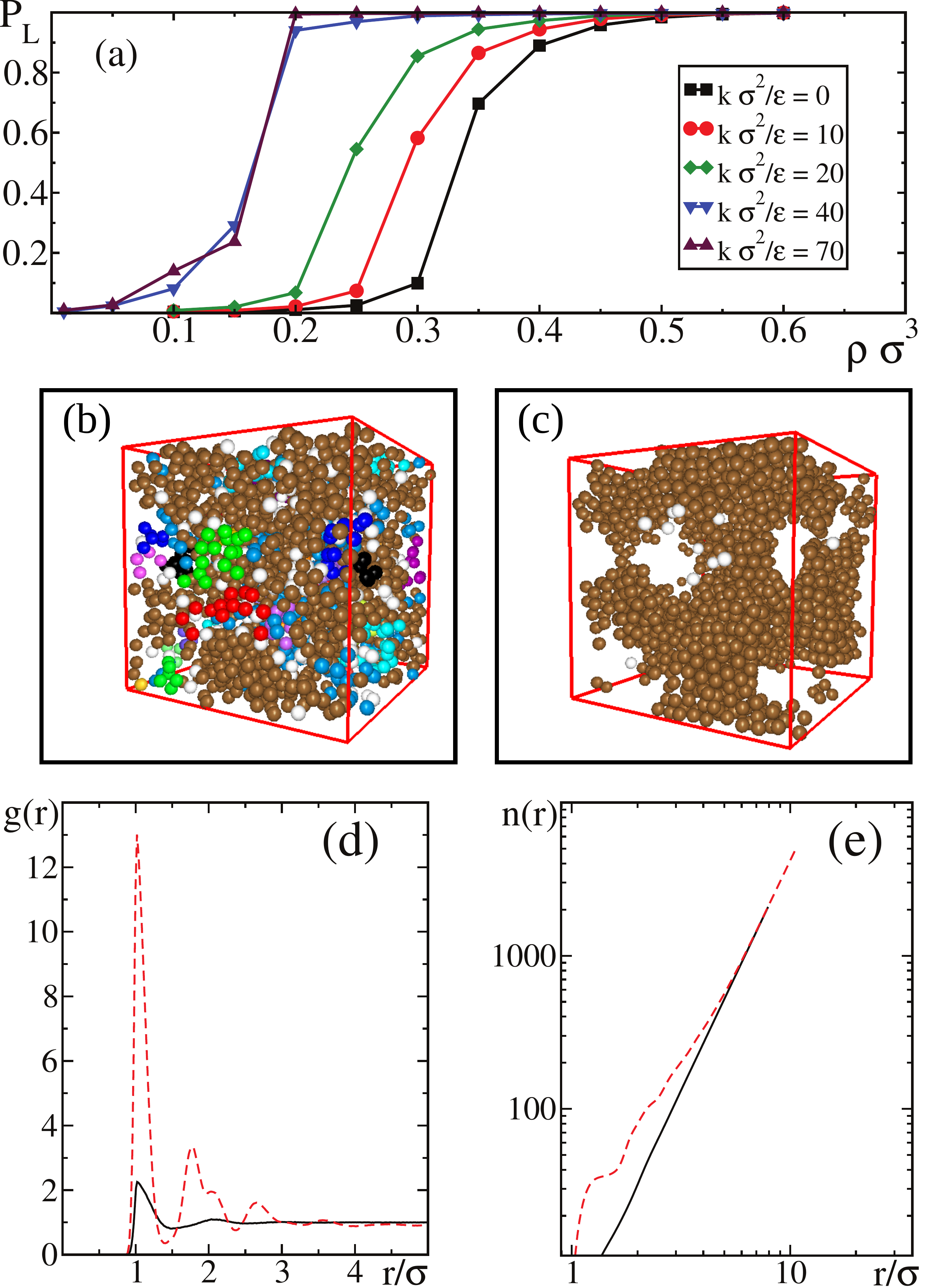}
\caption{Percolation transition: (a) Probability that a particle belongs to the largest cluster, $P_L$, as a function of the particle density. The different colors represent different strengths of the hysteretic spring. (b) Simulation snapshot of a percolated system with $\rho/\sigma^3=0.3$ and $k\sigma^2/\epsilon=10$. The largest cluster is colored in brown while white particles are not part of any cluster. (c) Snapshot with $\rho/\sigma^3=0.3$ and $k\sigma^2/\epsilon=40$ where the percolating cluster shows a fractal dimension $<$ 3. The coloring is similar to (b). (d) Radial distribution function for parameters $\rho \sigma^3 =0.2$ and $k \sigma^2 /\epsilon =70$ (red dashed curve), and $\rho \sigma^3 =0.5$ and $k \sigma^2 /\epsilon =10$ (black solid curve), representing both percolated systems. (e)Cumulative sum, $n(r)$, of (d) in double-log representation.\label{fig:perc}}
\end{figure}

We further investigate the structural properties of the system by investigating the percolation transition and the fractal dimension of percolated systems. We are interested in the critical density $\rho_c$ above which 50\% of the particles in the system belong to one cluster~\cite{REF12}, and especially in the dependence of $\rho_c$ on the hysteretic bond strength $k$. A cluster is an ensemble of particles that are connected so that it is possible to reach any particle in the cluster by following a path of bonds, regardless of the starting particle. Fig.~\ref{fig:perc} (a) shows the results for $P_L$, which is the ratio $N_{\rm CL}/N$, with $N_{CL}$ being the number of particles in the biggest cluster, and $N$ the total number of particles, as a function of density. The colors indicate different bond strengths. Clearly the percolation threshold $\rho_c$ decreases as $k$ is increased. The reason is the magnitude of the attractive pair interaction that increases with $k$;  a particle bonded with a strong hysteretic spring to a cluster is more unlikely to break away from it, compared to system with smaller $k$. This suggests that strong bonding supports  increased cluster growth and  increased stability of the cluster over time. The latter means that strongly interacting particles form percolating clusters that are stable.

The snapshots in Fig.~\ref{fig:perc} (b) and (c) show the system at $t /\tau_B = 80$. The colors indicate different clusters, where brown is the largest cluster and white particles do not belong to any cluster. In (b) the system with the parameters $k \sigma^2 /\epsilon = 10$ and $\rho \sigma^3 = 0.3$ is not percolated, i.e.\ the largest cluster does not contain 50\% of the particles. In (c) the system is percolated. Almost all particles belong to the percolating cluster for $k \sigma^2 /\epsilon = 40$ and $\rho \sigma^3 = 0.3$, where the particles act sticky. The snapshots reveal voids in the cluster, and therefore suggest a fractal dimension of the percolating cluster of $d_f < 3$.

In order to characterize the fractal structure, we calculate the cumulative sum of the radial distribution function $g(r)=G_{\rm dist}(r,t=0)$,
\begin{align}
n(r) = 4 \pi \rho \int_0^r r'^2 g(r') dr'~.
\end{align} 
It can be shown that $n(r)$ is related to the distance by a power law above a certain decay length 
\begin{align}
n(r) \propto r^{d_f},
\end{align}
with $d_f$ being the fractal dimension~\cite{vicsek}. The result is shown in Fig~\ref{fig:perc} (e), while in Fig~\ref{fig:perc} (d) the corresponding result for $g(r)$ is displayed. The black curve, where $\rho \sigma^3=0.5$ and $k \sigma^2 /\epsilon = 10$, shows a percolated system, where the fractal dimension is $d_f=3$. In the double-log plot of $n(r)$ this manifests itself by a straight line with slope 3. The red curve represents a percolated system with  $\rho \sigma^3=0.2$ and $k \sigma^2 /\epsilon = 70$. The percolating cluster has a fractal dimension of $d_f=2.31$, which is the slope of the red curve in Fig.~\ref{fig:perc} (e) when it starts to asymptotically approach the black line, around $3\lesssim r /\sigma \lesssim5$. Percolating clusters can be only found for systems with strong bonding, i.e. $k \sigma^2 / \epsilon=40$ and 70. With decreasing density, $d_f$ decreases. These values of the fractal dimension are consistent with fractal dimensions found in other colloidal systems~\cite{Poon:1995ts,REF0,fortini:pickering}, at intermediate densities and interaction strengths~\footnote{In the limit of low densities and large interaction strengths the system will reach a fractal dimension $d_f \simeq 1.7$, typical of systems formed from diffusion-limited cluster aggregation (DLCA)}.
The relative error of $d_f$ is rather large and can be estimated to be 15\%. The reason is that it is not always clear how to estimate the decay length from the graphical representation. Another error source is the fitting of a line to the relevant part of $n(r)$.

\subsection{van Hove Correlation function}
\label{sec:hove}

\begin{figure}
\centering
\includegraphics[width=8cm]{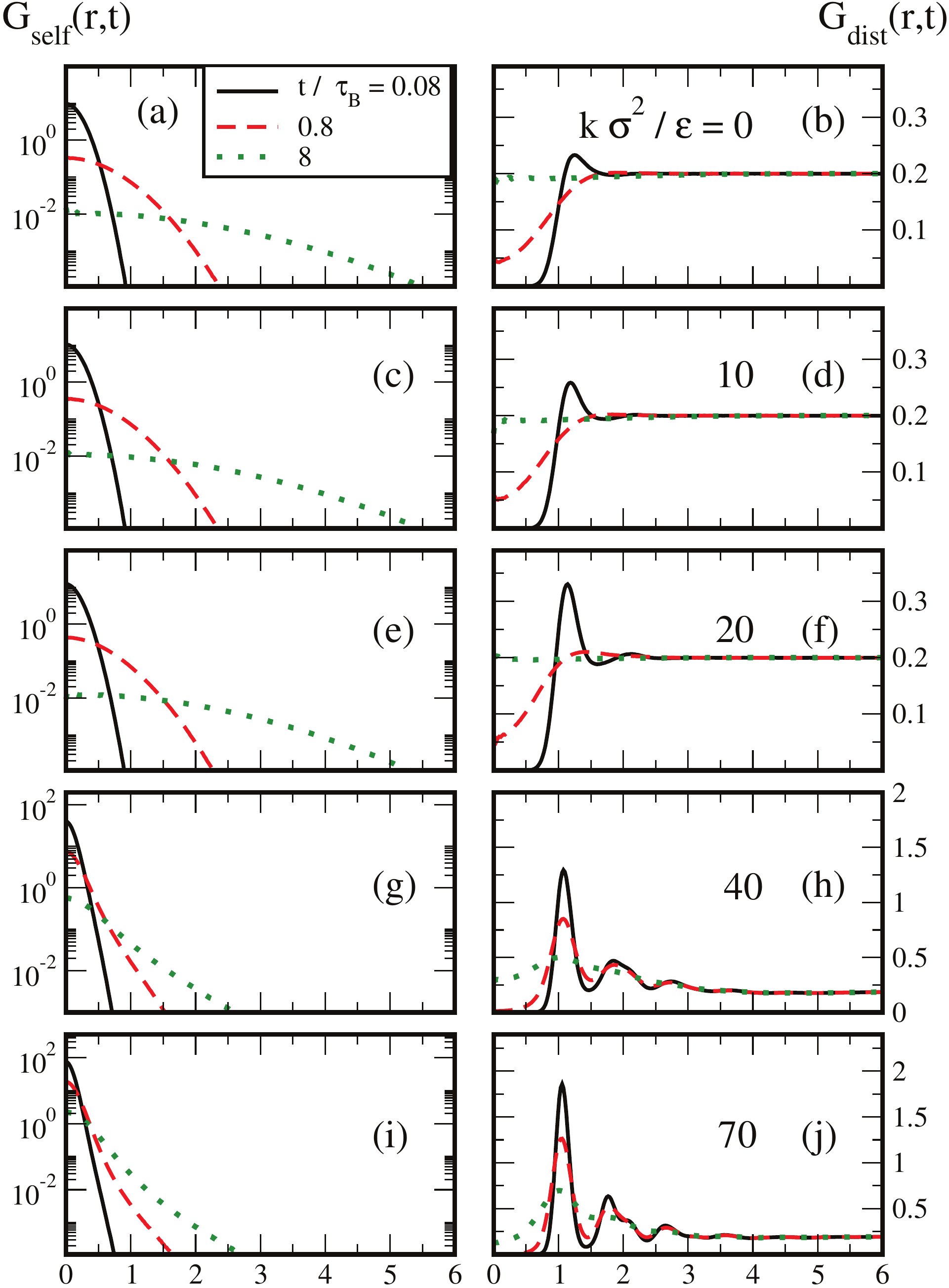}
\caption{The van Hove correlation function for $\rho \sigma^3=0.2$: The left column shows the self part of the van Hove function, the right column shows the distinct part at times $t/\tau_B = 0.08$  (black solid line),  $t/\tau_B=0.8$ (red dashed line) and $t/\tau_B=8$ (green dotted line). (a), (b) $k\sigma^2/\epsilon=0$, 
(c), (d) $k\sigma^2/\epsilon=10$, 
(e), (f) $k\sigma^2/\epsilon=20$,
(g), (h) $k\sigma^2/\epsilon=40$,
(i), (j) $k\sigma^2/\epsilon=70$.\label{fig:hove_02}}
\end{figure}

In Fig.~\ref{fig:hove_02} we show the results for the van Hove function for the density $\rho \sigma^3 = 0.2$. The left column shows the self-part $G_{\rm self}$ in semi-logarithmic representation, while the right column shows $G_{\rm dist}$ on a linear scale. The different colors and line styles indicate the different correlation times, where black solid is $t/\tau_B =0.08$, red dashed is $t/\tau_B =0.8$ and green dotted is $t/\tau_B =8$. In Fig.~\ref{fig:hove_02} in the first row $k \sigma^2 /\epsilon = 0$ (panels (a) and (b)), in the second row $k \sigma^2 /\epsilon = 10$, in the third row $k \sigma^2 /\epsilon = 20$, and in the last row $k \sigma^2 /\epsilon = 70$.

For increasing $k$ we observe an increase of the maximum height  of the self-part $G_{\rm self}$, as well as a decrease of its width (faster decay of the self-part of the correlation function). 
The reason is that at high $k$ the particles are more strongly bonded and they have a reduced mobility. 
Up to $k \sigma^2 /\epsilon = 20$ the shape of $G_{\rm self}$ is still a Gaussian, as expected for fluid systems~\cite{hansen:mcdonald}. If $k$ is increased to $k \sigma^2 /\epsilon = 40$ and beyond the shape of the self-part changes. This indicates the transition from a fluid to a network behavior. The deviation is quantified in more detail in Sec.~\ref{sec:ngauss}. 
We point out that deviations of the self part of the van Hove function from a Gaussian behavior correspond to the presence of the $\alpha$ and $\beta$ relaxation processes in the self intermediate scattering function~\cite{thevanhove}.

After the transition, the maximum of $G_{\rm self}$  increases by about two orders of magnitude and a fast decay of the correlation function, compared to fluid systems, indicates the presence of highly immobile particles in this region. 
Increasing the density has a similar effect on $G_{\rm self}$ as increasing $k$, though the reasons are different. The increase in $\rho$ leads to an increase of the number of collisions between particles, which also reduces their mobility. 
In comparable work, where only the density of a hard-sphere suspension is increased~\cite{thevanhove}, no crossover from a Gaussian shape was found, suggesting that the reduction of mobility is due to the hysteretic bonding drives this effect.

The distinct part of the van Hove function shows an increase of the height of the first peak for $t /\tau_B=0.08$, as $k$ is increased. The probability of finding a particle in the first correlation shell is increased, as $k$ is increased.  For $t /\tau_B = 0.8$ the peaks start to disappear and for $t /\tau_B=8$ $G_{\rm dist}$ is a constant in the fluid regime. For $k \sigma^2 /\epsilon \geq 40$, $G_{\rm dist}$ shows many oscillations, which only decrease in their amplitude, but do not vanish completely over time. 
This refers to a shell-like local structure, which is moderately stable over time, representing an arrested system. 
The dependency on density is comparable to that on $k$, but plays a more minor role. The reasons are similar to the ones given above. However, the transition to an arrested system is not observed, when only the density is increased.

\begin{figure}[htb]
\centering
\includegraphics[width=8cm]{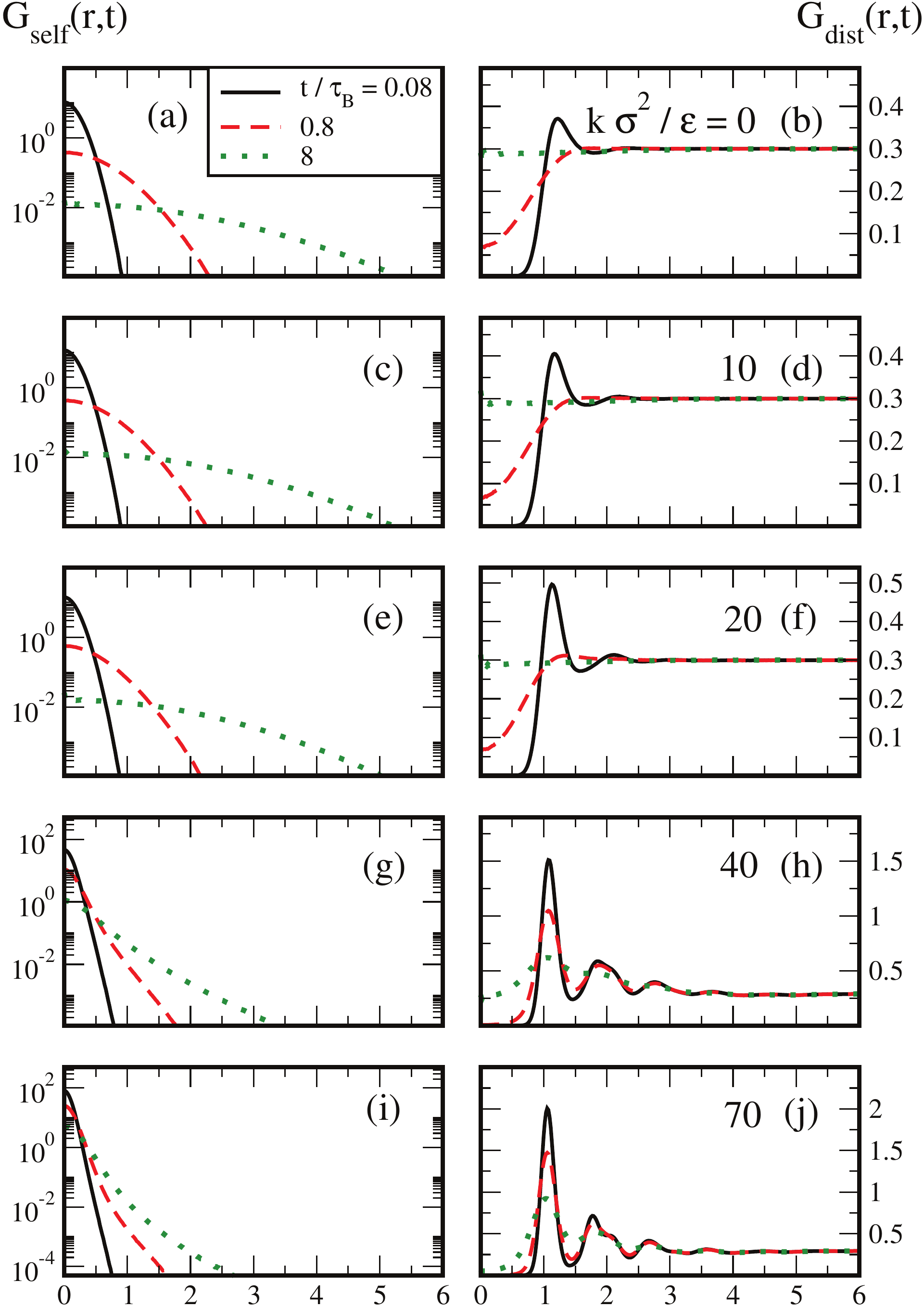}
\caption{ Same Fig.~\ref{fig:hove_02} but for  $\rho \sigma^3=0.3$.\label{fig:hove_03}}
\end{figure}

\begin{figure}[htb]
\centering
\includegraphics[width=8cm]{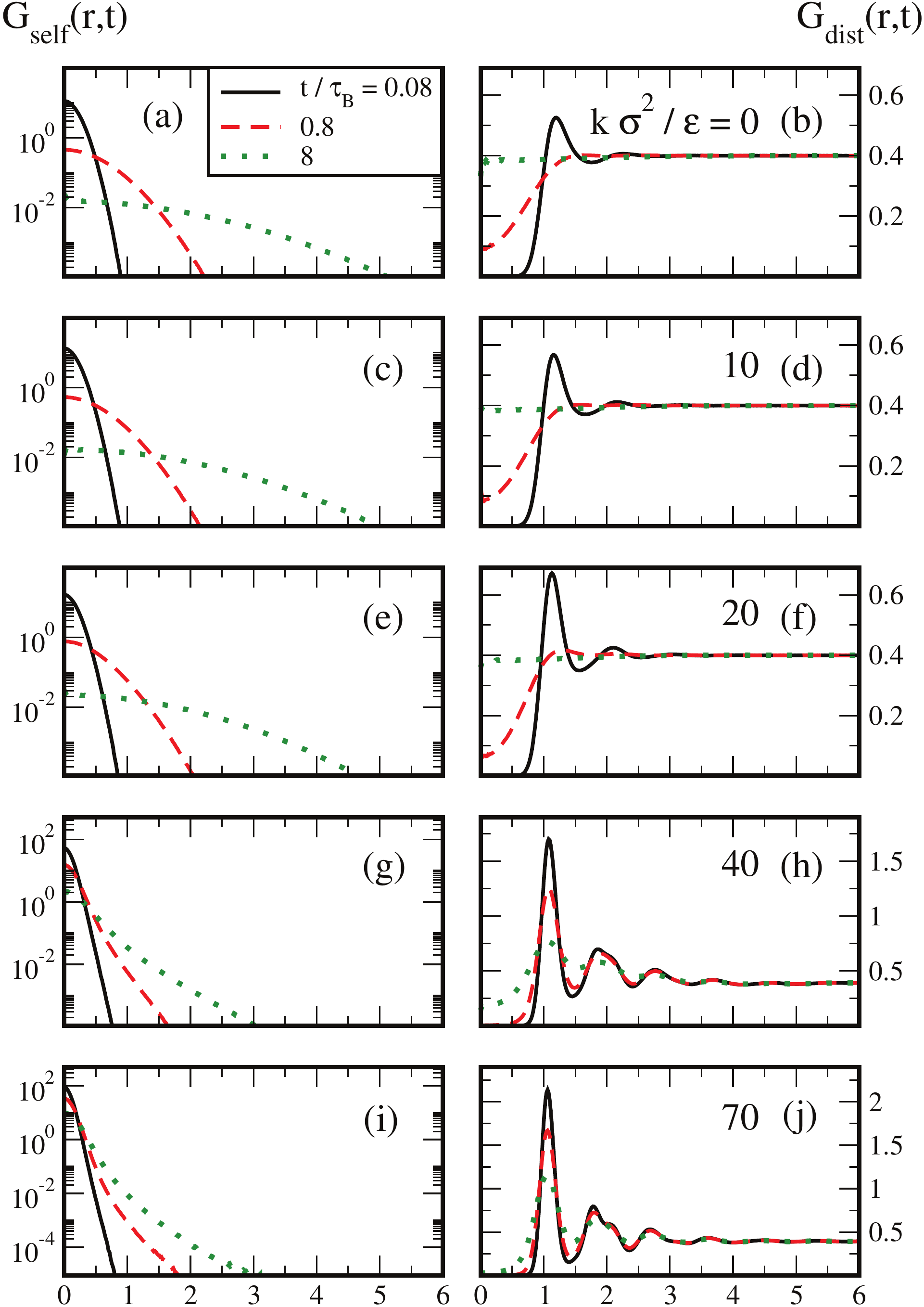}
\caption{Same as Fig.~\ref{fig:hove_03}, but $\rho \sigma^3=0.4$.\label{fig:hove_04}}
\end{figure}

Figure~\ref{fig:hove_03} shows the results for $G(r,t)$ for $\rho \sigma^3 =0.3$. The configuration of the panels and the color code are the same as in Fig.~\ref{fig:hove_02}. The observations are comparable to those for $\rho \sigma^3 =0.2$ and again we find a deviation from a Gaussian shape for $G_{\rm self}$ for $k \sigma^2 /\epsilon =40$. 
This indicates that the the transition from a fluid to a transient network occurs around $k \sigma^2 /\epsilon \simeq 40$.
This conclusion is supported by Fig.~\ref{fig:hove_04} where the van Hove function for $\rho \sigma^3 = 0.4$ is shown. The configuration of the panels is similar to Fig.~\ref{fig:hove_03}. As in the previous case, the deviation from a Gaussian in $G_{\rm self}$ and the arrested oscillations in $G_{\rm dist}$ are observed.

This results show that the model allows the tuning of the system's behavior from that of a fluid to that of a static network. The crossover is characterized by a transient network behavior. The system shows fluid-like dynamics with a reduced mobility of the particles, when bonds with finite strength are present.

\subsection{Non-Gaussian parameter}
\label{sec:ngauss}

\begin{figure}
\centering
\includegraphics[width=8cm]{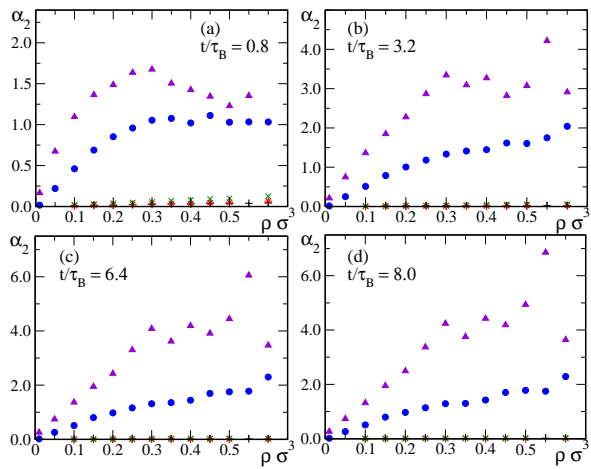}
\caption{Parameter $\alpha_2$ as a function of density: The panels show different correlation times, while the colors and symbol style refer to different bond strengths. In (a) the correlation time is $\tau/\tau_B=0.8$, in (b) 3.2, in (c) 6.4 and in (d) 8. The color and symbol code is the same for all times, with black crosses for $k \sigma^2 / \epsilon = 0$, red stars for $k \sigma^2 / \epsilon = 10$, green crosses for 20, blue diamonds for 40 and purple triangles for $k \sigma^2 / \epsilon = 70$. \label{fig:alpha2}}
\end{figure}

We next quantify the deviation of the self van Hove function from a fluid Gaussian behavior, by means of the parameter $\alpha_2$, defined in Eq.~(\ref{eq:ngauss}).
The results for $\alpha_2$ are shown in Fig.~\ref{fig:alpha2} as a function of density and for different correlation times  $t/\tau_B$ = 0.8, 3.2, 6.4, and 8. The color and symbol code refers to the bond strength.
We observe that $\alpha_2<0.1$ for all systems with $k \sigma^2 /\epsilon < 40$, regardless of density and time. Hence these systems can be regarded as fluid in the limit of the statistical error. For $k \sigma^2 /\epsilon=40$ the accordance of $G_{\rm self}$ with a Gaussian is quite good for $\rho \sigma^3 = 0.1$ for all times, but decreases rapidly as $\rho$ is increased until $\rho \sigma^3 \approx 0.3$. 
Above this density the increase of $\alpha_2$ slows down and for $t /\tau_B =0.8$, $\alpha_2$ almost saturates. The same behavior occurs for $k \sigma^2 /\epsilon=70$, but the non-Gaussian parameter has always a higher value than for $k \sigma^2 /\epsilon=40$. The saturation can be explained by the rate of collisions in the system, as a collision results in strong bonding for $k \sigma^2 /\epsilon=40$ and 70. 
In systems with $\rho \sigma^3 < 0.3$ collision events are rarer than in denser systems, indicating that more particles remain mobile because they diffusive freely between the collision events. 
The second moment of $G_{\rm self}$, which determines its width, decreases as time passes, and $k$ and $\rho$ are increased (see e.g.\ Fig.~\ref{fig:hove_04}). Moreover the fourth moment, the kurtosis, decreases too (see also\ Fig.~\ref{fig:hove_04}), but more rapidly than the second moment, because in total $\alpha_2$ increases. 
Therefore $\alpha_2$ quantifies the immobility of the particles compared to a fluid. We observe that $\alpha_2$ increases gradually when $\rho$ is increased. 
By varying $k$ and $\rho$, we can tune the system in a way that the deviation from a Gaussian of $G_{\rm self}$, $\alpha_2$ covers the dynamics of the system from fluid to fully static.

\section{Conclusion}
\label{sec:conclusion}

In conclusion, we have shown that the proposed model of hysteretic bond formation displays a variety of properties that are consistent with network formation. Depending on the strength of the bonding springs, we observe a crossover from transient network formation to an arrested quasi static network. We have used the two-body time-dependent (van Hove) correlation function to characterize the dynamic structure. A clear crossover from a fluid behavior at low spring constants to an arrested liquid-like structure at high spring constants and high densities is observed. This manifests itself in a clearly non-Gaussian shape of the self part and an increased correlation length and time in the distinct part for high spring constants. Moreover the crossover is quantized by the non-Gaussian parameter $\alpha_2$, which allows a more precise study of the crossover in the parameter range.

Moreover, we have find that the mobility of the particles can be tuned in the fluid regime by variation of the bond strength and the density. Our model can describe loose transient networks, where the rate of bonding and annihilation of bonds is high, as well as a strongly interacting network, in which new bonds last over a long period of time. This is supported by our statistical analysis of the bond forming and vanishing process. 

In future work, it would be very interesting to complement our simulation work by a theoretical approach that would describe network formation in fluids. One possible candidate for such a theory is the recent power functional approach, where the dynamics of a Brownian many-body systems is obtained from a variational principle on the one-body level~\cite{PFT}. Generalizing this approach in order to introduce the hysteretic bond formation process constitutes an interesting research task for the future.

The results presented in this paper pave the way to the analysis of the behavior of transient network formation with varying hysteretic behavior. 
The model allows one to change the critical parameters of the hysteretic interaction and evaluate the effects on network formation.
Hence, one is able to identify the signature behavior of hysteretic systems.  Known exampled of these systems, such as those governed by capillary forces, could also provide experimental confirmation.
Discovering non-obvious hysteretic behavior could be of importance for characterizing the network formation behavior of polymers or polymer particles, such as those used in the paint and coating industries ~\cite{Keddie:2010ta}.

\begin{acknowledgments}
We thank Joseph M. Brader and Thomas M. Fischer for helpful discussions.
PK acknowledges the Elitenetzwerk Bayern (ENB) for partial support.
\end{acknowledgments}


\begin{thebibliography}{48}%
\makeatletter
\providecommand \@ifxundefined [1]{%
 \@ifx{#1\undefined}
}%
\providecommand \@ifnum [1]{%
 \ifnum #1\expandafter \@firstoftwo
 \else \expandafter \@secondoftwo
 \fi
}%
\providecommand \@ifx [1]{%
 \ifx #1\expandafter \@firstoftwo
 \else \expandafter \@secondoftwo
 \fi
}%
\providecommand \natexlab [1]{#1}%
\providecommand \enquote  [1]{``#1''}%
\providecommand \bibnamefont  [1]{#1}%
\providecommand \bibfnamefont [1]{#1}%
\providecommand \citenamefont [1]{#1}%
\providecommand \href@noop [0]{\@secondoftwo}%
\providecommand \href [0]{\begingroup \@sanitize@url \@href}%
\providecommand \@href[1]{\@@startlink{#1}\@@href}%
\providecommand \@@href[1]{\endgroup#1\@@endlink}%
\providecommand \@sanitize@url [0]{\catcode `\\12\catcode `\$12\catcode
  `\&12\catcode `\#12\catcode `\^12\catcode `\_12\catcode `\%12\relax}%
\providecommand \@@startlink[1]{}%
\providecommand \@@endlink[0]{}%
\providecommand \url  [0]{\begingroup\@sanitize@url \@url }%
\providecommand \@url [1]{\endgroup\@href {#1}{\urlprefix }}%
\providecommand \urlprefix  [0]{URL }%
\providecommand \Eprint [0]{\href }%
\providecommand \doibase [0]{http://dx.doi.org/}%
\providecommand \selectlanguage [0]{\@gobble}%
\providecommand \bibinfo  [0]{\@secondoftwo}%
\providecommand \bibfield  [0]{\@secondoftwo}%
\providecommand \translation [1]{[#1]}%
\providecommand \BibitemOpen [0]{}%
\providecommand \bibitemStop [0]{}%
\providecommand \bibitemNoStop [0]{.\EOS\space}%
\providecommand \EOS [0]{\spacefactor3000\relax}%
\providecommand \BibitemShut  [1]{\csname bibitem#1\endcsname}%
\let\auto@bib@innerbib\@empty
\bibitem [{\citenamefont {Zaccarelli}(2007)}]{REF0}%
  \BibitemOpen
  \bibfield  {author} {\bibinfo {author} {\bibfnamefont {E.}~\bibnamefont
  {Zaccarelli}},\ }\href {http://stacks.iop.org/0953-8984/19/i=32/a=323101}
  {\bibfield  {journal} {\bibinfo  {journal} {J. Phy.: Cond. Mat.}\ }\textbf
  {\bibinfo {volume} {19}},\ \bibinfo {pages} {323101} (\bibinfo {year}
  {2007})}\BibitemShut {NoStop}%
\bibitem [{\citenamefont {Puertas}\ \emph {et~al.}(2004)\citenamefont
  {Puertas}, \citenamefont {Fuchs},\ and\ \citenamefont {Cates}}]{Puertes}%
  \BibitemOpen
  \bibfield  {author} {\bibinfo {author} {\bibfnamefont {A.~M.}\ \bibnamefont
  {Puertas}}, \bibinfo {author} {\bibfnamefont {M.}~\bibnamefont {Fuchs}}, \
  and\ \bibinfo {author} {\bibfnamefont {M.~E.}\ \bibnamefont {Cates}},\ }\href
  {\doibase 10.1063/1.1768936} {\bibfield  {journal} {\bibinfo  {journal} {J.
  Chem. Phys.}\ }\textbf {\bibinfo {volume} {121}},\ \bibinfo {pages} {2813}
  (\bibinfo {year} {2004})}\BibitemShut {NoStop}%
\bibitem [{\citenamefont {Lu}\ \emph {et~al.}(2008)\citenamefont {Lu},
  \citenamefont {Zaccarelli}, \citenamefont {Ciulla}, \citenamefont
  {Schofield}, \citenamefont {Sciortino},\ and\ \citenamefont
  {Weitz}}]{Nature}%
  \BibitemOpen
  \bibfield  {author} {\bibinfo {author} {\bibfnamefont {P.~J.}\ \bibnamefont
  {Lu}}, \bibinfo {author} {\bibfnamefont {E.}~\bibnamefont {Zaccarelli}},
  \bibinfo {author} {\bibfnamefont {F.}~\bibnamefont {Ciulla}}, \bibinfo
  {author} {\bibfnamefont {A.~B.}\ \bibnamefont {Schofield}}, \bibinfo {author}
  {\bibfnamefont {F.}~\bibnamefont {Sciortino}}, \ and\ \bibinfo {author}
  {\bibfnamefont {D.~A.}\ \bibnamefont {Weitz}},\ }\href {\doibase
  10.1038/nature06931} {\bibfield  {journal} {\bibinfo  {journal} {Nature}\
  }\textbf {\bibinfo {volume} {453}},\ \bibinfo {pages} {499} (\bibinfo {year}
  {2008})}\BibitemShut {NoStop}%
\bibitem [{\citenamefont {Green}\ and\ \citenamefont {Tobolsky}(1946)}]{Green}%
  \BibitemOpen
  \bibfield  {author} {\bibinfo {author} {\bibfnamefont {M.~S.}\ \bibnamefont
  {Green}}\ and\ \bibinfo {author} {\bibfnamefont {A.~V.}\ \bibnamefont
  {Tobolsky}},\ }\href {\doibase 10.1063/1.1724109} {\bibfield  {journal}
  {\bibinfo  {journal} {J. Chem. Phys.}\ }\textbf {\bibinfo {volume} {14}},\
  \bibinfo {pages} {80} (\bibinfo {year} {1946})}\BibitemShut {NoStop}%
\bibitem [{\citenamefont {Blair}\ and\ \citenamefont {Kudrolli}(2003)}]{Blair}%
  \BibitemOpen
  \bibfield  {author} {\bibinfo {author} {\bibfnamefont {D.~L.}\ \bibnamefont
  {Blair}}\ and\ \bibinfo {author} {\bibfnamefont {A.}~\bibnamefont
  {Kudrolli}},\ }\href {\doibase 10.1103/PhysRevE.67.021302} {\bibfield
  {journal} {\bibinfo  {journal} {Phys. Rev. E}\ }\textbf {\bibinfo {volume}
  {67}},\ \bibinfo {pages} {021302} (\bibinfo {year} {2003})}\BibitemShut
  {NoStop}%
\bibitem [{\citenamefont {Snoeijer}\ \emph {et~al.}(2004)\citenamefont
  {Snoeijer}, \citenamefont {Vlugt}, \citenamefont {van Hecke},\ and\
  \citenamefont {van Saarloos}}]{Snoeijer}%
  \BibitemOpen
  \bibfield  {author} {\bibinfo {author} {\bibfnamefont {J.~H.}\ \bibnamefont
  {Snoeijer}}, \bibinfo {author} {\bibfnamefont {T.~J.~H.}\ \bibnamefont
  {Vlugt}}, \bibinfo {author} {\bibfnamefont {M.}~\bibnamefont {van Hecke}}, \
  and\ \bibinfo {author} {\bibfnamefont {W.}~\bibnamefont {van Saarloos}},\
  }\href {\doibase 10.1103/PhysRevLett.92.054302} {\bibfield  {journal}
  {\bibinfo  {journal} {Phys. Rev. Lett.}\ }\textbf {\bibinfo {volume} {92}},\
  \bibinfo {pages} {054302} (\bibinfo {year} {2004})}\BibitemShut {NoStop}%
\bibitem [{\citenamefont {Utter}\ and\ \citenamefont
  {Behringer}(2004)}]{Utter}%
  \BibitemOpen
  \bibfield  {author} {\bibinfo {author} {\bibfnamefont {B.}~\bibnamefont
  {Utter}}\ and\ \bibinfo {author} {\bibfnamefont {R.}~\bibnamefont
  {Behringer}},\ }\href {\doibase 10.1140/epje/i2004-10022-4} {\bibfield
  {journal} {\bibinfo  {journal} {Euro. Phys. J. E}\ }\textbf {\bibinfo
  {volume} {14}},\ \bibinfo {pages} {373} (\bibinfo {year} {2004})}\BibitemShut
  {NoStop}%
\bibitem [{\citenamefont {Zucker}\ and\ \citenamefont {Regehr}(2002)}]{Zucker}%
  \BibitemOpen
  \bibfield  {author} {\bibinfo {author} {\bibfnamefont {R.~S.}\ \bibnamefont
  {Zucker}}\ and\ \bibinfo {author} {\bibfnamefont {W.~G.}\ \bibnamefont
  {Regehr}},\ }\href {\doibase 10.1146/annurev.physiol.64.092501.114547}
  {\bibfield  {journal} {\bibinfo  {journal} {Ann. Rev. Physiol.}\ }\textbf
  {\bibinfo {volume} {64}},\ \bibinfo {pages} {355} (\bibinfo {year}
  {2002})}\BibitemShut {NoStop}%
\bibitem [{\citenamefont {Tanaka}\ and\ \citenamefont
  {Edwards}(1992{\natexlab{a}})}]{REF1}%
  \BibitemOpen
  \bibfield  {author} {\bibinfo {author} {\bibfnamefont {F.}~\bibnamefont
  {Tanaka}}\ and\ \bibinfo {author} {\bibfnamefont {S.~F.}\ \bibnamefont
  {Edwards}},\ }\href {\doibase 10.1021/ma00031a024} {\bibfield  {journal}
  {\bibinfo  {journal} {Macromolecules}\ }\textbf {\bibinfo {volume} {25}},\
  \bibinfo {pages} {1516} (\bibinfo {year} {1992}{\natexlab{a}})}\BibitemShut
  {NoStop}%
\bibitem [{\citenamefont {Joshi}\ \emph {et~al.}(2000)\citenamefont {Joshi},
  \citenamefont {Lele},\ and\ \citenamefont {Mashelkar}}]{REF2}%
  \BibitemOpen
  \bibfield  {author} {\bibinfo {author} {\bibfnamefont {Y.~M.}\ \bibnamefont
  {Joshi}}, \bibinfo {author} {\bibfnamefont {A.~K.}\ \bibnamefont {Lele}}, \
  and\ \bibinfo {author} {\bibfnamefont {R.}~\bibnamefont {Mashelkar}},\ }\href
  {\doibase http://dx.doi.org/10.1016/S0377-0257(99)00046-4} {\bibfield
  {journal} {\bibinfo  {journal} {J. Non-Newt. Fluid Mech.}\ }\textbf {\bibinfo
  {volume} {89}},\ \bibinfo {pages} {303 } (\bibinfo {year}
  {2000})}\BibitemShut {NoStop}%
\bibitem [{\citenamefont {Jongschaap}\ \emph {et~al.}(2001)\citenamefont
  {Jongschaap}, \citenamefont {Wientjes}, \citenamefont {Duits},\ and\
  \citenamefont {Mellema}}]{REF3}%
  \BibitemOpen
  \bibfield  {author} {\bibinfo {author} {\bibfnamefont {R.~J.~J.}\
  \bibnamefont {Jongschaap}}, \bibinfo {author} {\bibfnamefont {R.~H.~W.}\
  \bibnamefont {Wientjes}}, \bibinfo {author} {\bibfnamefont {M.~H.~G.}\
  \bibnamefont {Duits}}, \ and\ \bibinfo {author} {\bibfnamefont
  {J.}~\bibnamefont {Mellema}},\ }\href {\doibase 10.1021/ma0001640} {\bibfield
   {journal} {\bibinfo  {journal} {Macromolecules}\ }\textbf {\bibinfo {volume}
  {34}},\ \bibinfo {pages} {1031} (\bibinfo {year} {2001})}\BibitemShut
  {NoStop}%
\bibitem [{\citenamefont {Linder}\ \emph {et~al.}(2011)\citenamefont {Linder},
  \citenamefont {Tkachuk},\ and\ \citenamefont {Miehe}}]{REF4}%
  \BibitemOpen
  \bibfield  {author} {\bibinfo {author} {\bibfnamefont {C.}~\bibnamefont
  {Linder}}, \bibinfo {author} {\bibfnamefont {M.}~\bibnamefont {Tkachuk}}, \
  and\ \bibinfo {author} {\bibfnamefont {C.}~\bibnamefont {Miehe}},\ }\href
  {\doibase http://dx.doi.org/10.1016/j.jmps.2011.05.005} {\bibfield  {journal}
  {\bibinfo  {journal} {J. Mech. Phys. Sol.}\ }\textbf {\bibinfo {volume}
  {59}},\ \bibinfo {pages} {2134 } (\bibinfo {year} {2011})}\BibitemShut
  {NoStop}%
\bibitem [{\citenamefont {Annable}\ \emph {et~al.}(1993)\citenamefont
  {Annable}, \citenamefont {Buscall}, \citenamefont {Ettelaie},\ and\
  \citenamefont {Whittlestone}}]{REF5}%
  \BibitemOpen
  \bibfield  {author} {\bibinfo {author} {\bibfnamefont {T.}~\bibnamefont
  {Annable}}, \bibinfo {author} {\bibfnamefont {R.}~\bibnamefont {Buscall}},
  \bibinfo {author} {\bibfnamefont {R.}~\bibnamefont {Ettelaie}}, \ and\
  \bibinfo {author} {\bibfnamefont {D.}~\bibnamefont {Whittlestone}},\ }\href
  {\doibase http://dx.doi.org/10.1122/1.550391} {\bibfield  {journal} {\bibinfo
   {journal} {J. Rheol.}\ }\textbf {\bibinfo {volume} {37}},\ \bibinfo {pages}
  {695} (\bibinfo {year} {1993})}\BibitemShut {NoStop}%
\bibitem [{\citenamefont {De~Gennes}(1979)}]{deGennes}%
  \BibitemOpen
  \bibfield  {author} {\bibinfo {author} {\bibfnamefont {P.~G.}\ \bibnamefont
  {De~Gennes}},\ }\href@noop {} {\emph {\bibinfo {title} {Scaling Concepts in
  Polymer Physics}}}\ (\bibinfo  {publisher} {Cornell University Press},\
  \bibinfo {year} {1979})\BibitemShut {NoStop}%
\bibitem [{\citenamefont {Michel}\ \emph {et~al.}(2000)\citenamefont {Michel},
  \citenamefont {Filali}, \citenamefont {Aznar}, \citenamefont {Porte},\ and\
  \citenamefont {Appell}}]{REF6}%
  \BibitemOpen
  \bibfield  {author} {\bibinfo {author} {\bibfnamefont {E.}~\bibnamefont
  {Michel}}, \bibinfo {author} {\bibfnamefont {M.}~\bibnamefont {Filali}},
  \bibinfo {author} {\bibfnamefont {R.}~\bibnamefont {Aznar}}, \bibinfo
  {author} {\bibfnamefont {G.}~\bibnamefont {Porte}}, \ and\ \bibinfo {author}
  {\bibfnamefont {J.}~\bibnamefont {Appell}},\ }\href {\doibase
  10.1021/la000317c} {\bibfield  {journal} {\bibinfo  {journal} {Langmuir}\
  }\textbf {\bibinfo {volume} {16}},\ \bibinfo {pages} {8702} (\bibinfo {year}
  {2000})}\BibitemShut {NoStop}%
\bibitem [{\citenamefont {Verhaegh}\ \emph {et~al.}(1999)\citenamefont
  {Verhaegh}, \citenamefont {Asnaghi},\ and\ \citenamefont
  {Lekkerkerker}}]{REF7}%
  \BibitemOpen
  \bibfield  {author} {\bibinfo {author} {\bibfnamefont {N.~A.~M.}\
  \bibnamefont {Verhaegh}}, \bibinfo {author} {\bibfnamefont {D.}~\bibnamefont
  {Asnaghi}}, \ and\ \bibinfo {author} {\bibfnamefont {H.~N.~W.}\ \bibnamefont
  {Lekkerkerker}},\ }\href {\doibase
  http://dx.doi.org/10.1016/S0378-4371(98)00420-8} {\bibfield  {journal}
  {\bibinfo  {journal} {Physica A}\ }\textbf {\bibinfo {volume} {264}},\
  \bibinfo {pages} {64 } (\bibinfo {year} {1999})}\BibitemShut {NoStop}%
\bibitem [{\citenamefont {Tanaka}\ \emph {et~al.}(2005)\citenamefont {Tanaka},
  \citenamefont {Nishikawa},\ and\ \citenamefont {Koyama}}]{REF8}%
  \BibitemOpen
  \bibfield  {author} {\bibinfo {author} {\bibfnamefont {H.}~\bibnamefont
  {Tanaka}}, \bibinfo {author} {\bibfnamefont {Y.}~\bibnamefont {Nishikawa}}, \
  and\ \bibinfo {author} {\bibfnamefont {T.}~\bibnamefont {Koyama}},\ }\href
  {http://stacks.iop.org/0953-8984/17/i=15/a=L02} {\bibfield  {journal}
  {\bibinfo  {journal} {J. Phy.: Cond. Mat.}\ }\textbf {\bibinfo {volume}
  {17}},\ \bibinfo {pages} {L143} (\bibinfo {year} {2005})}\BibitemShut
  {NoStop}%
\bibitem [{\citenamefont {Osterman}\ \emph {et~al.}(2009)\citenamefont
  {Osterman}, \citenamefont {Poberaj}, \citenamefont {Dobnikar}, \citenamefont
  {Frenkel}, \citenamefont {Ziherl},\ and\ \citenamefont
  {Babi\ifmmode~\acute{c}\else \'{c}\fi{}}}]{Osterman}%
  \BibitemOpen
  \bibfield  {author} {\bibinfo {author} {\bibfnamefont {N.}~\bibnamefont
  {Osterman}}, \bibinfo {author} {\bibfnamefont {I.}~\bibnamefont {Poberaj}},
  \bibinfo {author} {\bibfnamefont {J.}~\bibnamefont {Dobnikar}}, \bibinfo
  {author} {\bibfnamefont {D.}~\bibnamefont {Frenkel}}, \bibinfo {author}
  {\bibfnamefont {P.}~\bibnamefont {Ziherl}}, \ and\ \bibinfo {author}
  {\bibfnamefont {D.}~\bibnamefont {Babi\ifmmode~\acute{c}\else \'{c}\fi{}}},\
  }\href {\doibase 10.1103/PhysRevLett.103.228301} {\bibfield  {journal}
  {\bibinfo  {journal} {Phys. Rev. Lett.}\ }\textbf {\bibinfo {volume} {103}},\
  \bibinfo {pages} {228301} (\bibinfo {year} {2009})}\BibitemShut {NoStop}%
\bibitem{Maier}
F.~J. Maier and T.~M. Fischer, Soft Matter {\bf 12}, 614 (2016).   
\bibitem [{\citenamefont {Zilman}\ \emph {et~al.}(2003)\citenamefont {Zilman},
  \citenamefont {Tlusty},\ and\ \citenamefont {Safran}}]{Referee1}%
  \BibitemOpen
  \bibfield  {author} {\bibinfo {author} {\bibfnamefont {A.}~\bibnamefont
  {Zilman}}, \bibinfo {author} {\bibfnamefont {T.}~\bibnamefont {Tlusty}}, \
  and\ \bibinfo {author} {\bibfnamefont {S.~A.}\ \bibnamefont {Safran}},\
  }\href {http://stacks.iop.org/0953-8984/15/i=1/a=306} {\bibfield  {journal}
  {\bibinfo  {journal} {J. Phy.: Cond. Mat.}\ }\textbf {\bibinfo {volume}
  {15}},\ \bibinfo {pages} {S57} (\bibinfo {year} {2003})}\BibitemShut
  {NoStop}%
\bibitem [{\citenamefont {Blaak}\ \emph {et~al.}(2007)\citenamefont {Blaak},
  \citenamefont {Miller},\ and\ \citenamefont {Hansen}}]{Referee2}%
  \BibitemOpen
  \bibfield  {author} {\bibinfo {author} {\bibfnamefont {R.}~\bibnamefont
  {Blaak}}, \bibinfo {author} {\bibfnamefont {M.~A.}\ \bibnamefont {Miller}}, \
  and\ \bibinfo {author} {\bibfnamefont {J.-P.}\ \bibnamefont {Hansen}},\
  }\href {http://dx.doi.org/10.1209/0295-5075/78/26002} {\bibfield  {journal}
  {\bibinfo  {journal} {EPL}\ }\textbf {\bibinfo {volume} {78}},\ \bibinfo
  {pages} {26002} (\bibinfo {year} {2007})}\BibitemShut {NoStop}%
\bibitem [{\citenamefont {Miller}\ \emph {et~al.}(2009)\citenamefont {Miller},
  \citenamefont {Blaak}, \citenamefont {Lumb},\ and\ \citenamefont
  {Hansen}}]{Referee3}%
  \BibitemOpen
  \bibfield  {author} {\bibinfo {author} {\bibfnamefont {M.~A.}\ \bibnamefont
  {Miller}}, \bibinfo {author} {\bibfnamefont {R.}~\bibnamefont {Blaak}},
  \bibinfo {author} {\bibfnamefont {C.~N.}\ \bibnamefont {Lumb}}, \ and\
  \bibinfo {author} {\bibfnamefont {J.-P.}\ \bibnamefont {Hansen}},\ }\href
  {http://scitation.aip.org/content/aip/journal/jcp/130/11/10.1063/1.3089620}
  {\bibfield  {journal} {\bibinfo  {journal} {J. Chem. Phy.}\ }\textbf
  {\bibinfo {volume} {130}},\ \bibinfo {eid} {114507} (\bibinfo {year}
  {2009})}\BibitemShut {NoStop}%
\bibitem [{\citenamefont {H{\"u}tter}(2000)}]{Hutter}%
  \BibitemOpen
  \bibfield  {author} {\bibinfo {author} {\bibfnamefont {M.}~\bibnamefont
  {H{\"u}tter}},\ }\href {\doibase 10.1006/jcis.2000.7150} {\bibfield
  {journal} {\bibinfo  {journal} {J. Coll. Interf. Sci.}\ }\textbf {\bibinfo
  {volume} {231}},\ \bibinfo {pages} {337} (\bibinfo {year}
  {2000})}\BibitemShut {NoStop}%
\bibitem [{\citenamefont {Zhang}\ and\ \citenamefont {Glotzer}(2004)}]{REF9}%
  \BibitemOpen
  \bibfield  {author} {\bibinfo {author} {\bibfnamefont {Z.}~\bibnamefont
  {Zhang}}\ and\ \bibinfo {author} {\bibfnamefont {S.~C.}\ \bibnamefont
  {Glotzer}},\ }\href {\doibase 10.1021/nl0493500} {\bibfield  {journal}
  {\bibinfo  {journal} {Nano Letters}\ }\textbf {\bibinfo {volume} {4}},\
  \bibinfo {pages} {1407} (\bibinfo {year} {2004})}\BibitemShut {NoStop}%
\bibitem [{\citenamefont {Zhang}\ \emph {et~al.}(2005)\citenamefont {Zhang},
  \citenamefont {Keys}, \citenamefont {Chen},\ and\ \citenamefont
  {Glotzer}}]{REF10}%
  \BibitemOpen
  \bibfield  {author} {\bibinfo {author} {\bibfnamefont {Z.}~\bibnamefont
  {Zhang}}, \bibinfo {author} {\bibfnamefont {A.~S.}\ \bibnamefont {Keys}},
  \bibinfo {author} {\bibfnamefont {T.}~\bibnamefont {Chen}}, \ and\ \bibinfo
  {author} {\bibfnamefont {S.~C.}\ \bibnamefont {Glotzer}},\ }\href {\doibase
  10.1021/la0513611} {\bibfield  {journal} {\bibinfo  {journal} {Langmuir}\
  }\textbf {\bibinfo {volume} {21}},\ \bibinfo {pages} {11547} (\bibinfo {year}
  {2005})}\BibitemShut {NoStop}%
\bibitem [{\citenamefont {Bianchi}\ \emph {et~al.}(2006)\citenamefont
  {Bianchi}, \citenamefont {Largo}, \citenamefont {Tartaglia}, \citenamefont
  {Zaccarelli},\ and\ \citenamefont {Sciortino}}]{REF11}%
  \BibitemOpen
  \bibfield  {author} {\bibinfo {author} {\bibfnamefont {E.}~\bibnamefont
  {Bianchi}}, \bibinfo {author} {\bibfnamefont {J.}~\bibnamefont {Largo}},
  \bibinfo {author} {\bibfnamefont {P.}~\bibnamefont {Tartaglia}}, \bibinfo
  {author} {\bibfnamefont {E.}~\bibnamefont {Zaccarelli}}, \ and\ \bibinfo
  {author} {\bibfnamefont {F.}~\bibnamefont {Sciortino}},\ }\href {\doibase
  10.1103/PhysRevLett.97.168301} {\bibfield  {journal} {\bibinfo  {journal}
  {Phys. Rev. Lett.}\ }\textbf {\bibinfo {volume} {97}},\ \bibinfo {pages}
  {168301} (\bibinfo {year} {2006})}\BibitemShut {NoStop}%
\bibitem [{\citenamefont {de~las Heras}\ \emph
  {et~al.}(2011{\natexlab{a}})\citenamefont {de~las Heras}, \citenamefont
  {Tavares},\ and\ \citenamefont {Telo~da Gama}}]{dani_patchy}%
  \BibitemOpen
  \bibfield  {author} {\bibinfo {author} {\bibfnamefont {D.}~\bibnamefont
  {de~las Heras}}, \bibinfo {author} {\bibfnamefont {J.~M.}\ \bibnamefont
  {Tavares}}, \ and\ \bibinfo {author} {\bibfnamefont {M.~M.}\ \bibnamefont
  {Telo~da Gama}},\ }\href@noop {} {\bibfield  {journal} {\bibinfo  {journal}
  {J. Chem. Phys.}\ }\textbf {\bibinfo {volume} {134}} (\bibinfo {year}
  {2011}{\natexlab{a}})}\BibitemShut {NoStop}%
\bibitem [{\citenamefont {Dias}\ \emph {et~al.}(2014)\citenamefont {Dias},
  \citenamefont {Ara\'ujo},\ and\ \citenamefont {Telo~da Gama}}]{Dias}%
  \BibitemOpen
  \bibfield  {author} {\bibinfo {author} {\bibfnamefont {C.~S.}\ \bibnamefont
  {Dias}}, \bibinfo {author} {\bibfnamefont {N.~A.~M.}\ \bibnamefont
  {Ara\'ujo}}, \ and\ \bibinfo {author} {\bibfnamefont {M.~M.}\ \bibnamefont
  {Telo~da Gama}},\ }\href {\doibase 10.1103/PhysRevE.90.032302} {\bibfield
  {journal} {\bibinfo  {journal} {Phys. Rev. E}\ }\textbf {\bibinfo {volume}
  {90}},\ \bibinfo {pages} {032302} (\bibinfo {year} {2014})}\BibitemShut
  {NoStop}%
\bibitem [{\citenamefont {de~las Heras}\ \emph
  {et~al.}(2011{\natexlab{b}})\citenamefont {de~las Heras}, \citenamefont
  {Tavares},\ and\ \citenamefont {Telo~da Gama}}]{dani_network}%
  \BibitemOpen
  \bibfield  {author} {\bibinfo {author} {\bibfnamefont {D.}~\bibnamefont
  {de~las Heras}}, \bibinfo {author} {\bibfnamefont {J.~M.}\ \bibnamefont
  {Tavares}}, \ and\ \bibinfo {author} {\bibfnamefont {M.~M.}\ \bibnamefont
  {Telo~da Gama}},\ }\href@noop {} {\bibfield  {journal} {\bibinfo  {journal}
  {Soft Matter}\ }\textbf {\bibinfo {volume} {7}} (\bibinfo {year}
  {2011}{\natexlab{b}})}\BibitemShut {NoStop}%
\bibitem [{\citenamefont {Strobl}(2007)}]{Strobl}%
  \BibitemOpen
  \bibfield  {author} {\bibinfo {author} {\bibfnamefont {G.}~\bibnamefont
  {Strobl}},\ }\href@noop {} {\emph {\bibinfo {title} {The Physics of
  Polymers}}}\ (\bibinfo  {publisher} {Springer Berlin},\ \bibinfo {year}
  {2007})\BibitemShut {NoStop}%
\bibitem [{\citenamefont {Krinninger}\ \emph {et~al.}(2014)\citenamefont
  {Krinninger}, \citenamefont {Fischer},\ and\ \citenamefont
  {Fortini}}]{Krinninger}%
  \BibitemOpen
  \bibfield  {author} {\bibinfo {author} {\bibfnamefont {P.}~\bibnamefont
  {Krinninger}}, \bibinfo {author} {\bibfnamefont {A.}~\bibnamefont {Fischer}},
  \ and\ \bibinfo {author} {\bibfnamefont {A.}~\bibnamefont {Fortini}},\ }\href
  {\doibase 10.1103/PhysRevE.90.012201} {\bibfield  {journal} {\bibinfo
  {journal} {Phys. Rev. E}\ }\textbf {\bibinfo {volume} {90}},\ \bibinfo
  {pages} {012201} (\bibinfo {year} {2014})}\BibitemShut {NoStop}%
\bibitem [{\citenamefont {Herminghaus}(2005)}]{Herminghaus}%
  \BibitemOpen
  \bibfield  {author} {\bibinfo {author} {\bibfnamefont {S.}~\bibnamefont
  {Herminghaus}},\ }\href {\doibase 10.1080/00018730500167855} {\bibfield
  {journal} {\bibinfo  {journal} {Adv. Phys.}\ }\textbf {\bibinfo {volume}
  {54}},\ \bibinfo {pages} {221} (\bibinfo {year} {2005})}\BibitemShut
  {NoStop}%
\bibitem [{\citenamefont {van Hove}(1954)}]{vhove:vhove}%
  \BibitemOpen
  \bibfield  {author} {\bibinfo {author} {\bibfnamefont {L.}~\bibnamefont {van
  Hove}},\ }\href {\doibase 10.1103/PhysRev.95.249} {\bibfield  {journal}
  {\bibinfo  {journal} {Phys. Rev.}\ }\textbf {\bibinfo {volume} {95}},\
  \bibinfo {pages} {249} (\bibinfo {year} {1954})}\BibitemShut {NoStop}%
\bibitem [{\citenamefont {Hansen}\ and\ \citenamefont
  {McDonald}(2013)}]{hansen:mcdonald}%
  \BibitemOpen
  \bibfield  {author} {\bibinfo {author} {\bibfnamefont {J.~P.}\ \bibnamefont
  {Hansen}}\ and\ \bibinfo {author} {\bibfnamefont {I.~R.}\ \bibnamefont
  {McDonald}},\ }\href@noop {} {\emph {\bibinfo {title} {Theory of Simple
  Liquids}}}\ (\bibinfo  {publisher} {Elsevier Science},\ \bibinfo {year}
  {2013})\BibitemShut {NoStop}%
\bibitem [{\citenamefont {Brader}\ and\ \citenamefont {Schmidt}(2013)}]{NOZ1}%
  \BibitemOpen
  \bibfield  {author} {\bibinfo {author} {\bibfnamefont {J.~M.}\ \bibnamefont
  {Brader}}\ and\ \bibinfo {author} {\bibfnamefont {M.}~\bibnamefont
  {Schmidt}},\ }\href {\doibase http://dx.doi.org/10.1063/1.4820399} {\bibfield
   {journal} {\bibinfo  {journal} {J. Chem. Phys.}\ }\textbf {\bibinfo {volume}
  {139}},\ \bibinfo {eid} {104108} (\bibinfo {year} {2013})}\BibitemShut
  {NoStop}%
\bibitem [{\citenamefont {Brader}\ and\ \citenamefont {Schmidt}(2014)}]{NOZ2}%
  \BibitemOpen
  \bibfield  {author} {\bibinfo {author} {\bibfnamefont {J.~M.}\ \bibnamefont
  {Brader}}\ and\ \bibinfo {author} {\bibfnamefont {M.}~\bibnamefont
  {Schmidt}},\ }\href {\doibase http://dx.doi.org/10.1063/1.4861041} {\bibfield
   {journal} {\bibinfo  {journal} {J. Chem. Phys.}\ }\textbf {\bibinfo {volume}
  {140}},\ \bibinfo {eid} {034104} (\bibinfo {year} {2014})}\BibitemShut
  {NoStop}%
\bibitem [{\citenamefont {Schmidt}\ and\ \citenamefont {Brader}(2013)}]{PFT}%
  \BibitemOpen
  \bibfield  {author} {\bibinfo {author} {\bibfnamefont {M.}~\bibnamefont
  {Schmidt}}\ and\ \bibinfo {author} {\bibfnamefont {J.~M.}\ \bibnamefont
  {Brader}},\ }\href {\doibase http://dx.doi.org/10.1063/1.4807586} {\bibfield
  {journal} {\bibinfo  {journal} {J. Chem. Phys.}\ }\textbf {\bibinfo {volume}
  {138}},\ \bibinfo {eid} {214101} (\bibinfo {year} {2013})}\BibitemShut
  {NoStop}%
\bibitem [{\citenamefont {Hopkins}\ \emph {et~al.}(2010)\citenamefont
  {Hopkins}, \citenamefont {Fortini}, \citenamefont {Archer},\ and\
  \citenamefont {Schmidt}}]{thevanhove}%
  \BibitemOpen
  \bibfield  {author} {\bibinfo {author} {\bibfnamefont {P.}~\bibnamefont
  {Hopkins}}, \bibinfo {author} {\bibfnamefont {A.}~\bibnamefont {Fortini}},
  \bibinfo {author} {\bibfnamefont {A.~J.}\ \bibnamefont {Archer}}, \ and\
  \bibinfo {author} {\bibfnamefont {M.}~\bibnamefont {Schmidt}},\ }\href
  {\doibase 10.1063/1.3511719} {\bibfield  {journal} {\bibinfo  {journal} {J.
  Chem. Phys.}\ }\textbf {\bibinfo {volume} {133}},\ \bibinfo {eid} {224505}
  (\bibinfo {year} {2010})}\BibitemShut {NoStop}%
\bibitem [{\citenamefont {Archer}\ \emph {et~al.}(2007)\citenamefont {Archer},
  \citenamefont {Hopkins},\ and\ \citenamefont {Schmidt}}]{ajarcher}%
  \BibitemOpen
  \bibfield  {author} {\bibinfo {author} {\bibfnamefont {A.~J.}\ \bibnamefont
  {Archer}}, \bibinfo {author} {\bibfnamefont {P.}~\bibnamefont {Hopkins}}, \
  and\ \bibinfo {author} {\bibfnamefont {M.}~\bibnamefont {Schmidt}},\ }\href
  {\doibase 10.1103/PhysRevE.75.040501} {\bibfield  {journal} {\bibinfo
  {journal} {Phys. Rev. E}\ }\textbf {\bibinfo {volume} {75}},\ \bibinfo
  {pages} {040501} (\bibinfo {year} {2007})}\BibitemShut {NoStop}%
\bibitem [{\citenamefont {Brader}\ and\ \citenamefont {Schmidt}(2015)}]{TPL}%
  \BibitemOpen
  \bibfield  {author} {\bibinfo {author} {\bibfnamefont {J.~M.}\ \bibnamefont
  {Brader}}\ and\ \bibinfo {author} {\bibfnamefont {M.}~\bibnamefont
  {Schmidt}},\ }\href {http://stacks.iop.org/0953-8984/27/i=19/a=194106}
  {\bibfield  {journal} {\bibinfo  {journal} {J. Phys.: Cond. Mat.}\ }\textbf
  {\bibinfo {volume} {27}},\ \bibinfo {pages} {194106} (\bibinfo {year}
  {2015})}\BibitemShut {NoStop}%
\bibitem [{\citenamefont {Donati}\ \emph {et~al.}(1999)\citenamefont {Donati},
  \citenamefont {Glotzer}, \citenamefont {Poole}, \citenamefont {Kob},\ and\
  \citenamefont {Plimpton}}]{ngauss:kob}%
  \BibitemOpen
  \bibfield  {author} {\bibinfo {author} {\bibfnamefont {C.}~\bibnamefont
  {Donati}}, \bibinfo {author} {\bibfnamefont {S.~C.}\ \bibnamefont {Glotzer}},
  \bibinfo {author} {\bibfnamefont {P.~H.}\ \bibnamefont {Poole}}, \bibinfo
  {author} {\bibfnamefont {W.}~\bibnamefont {Kob}}, \ and\ \bibinfo {author}
  {\bibfnamefont {S.~J.}\ \bibnamefont {Plimpton}},\ }\href {\doibase
  10.1103/PhysRevE.60.3107} {\bibfield  {journal} {\bibinfo  {journal} {Phys.
  Rev. E}\ }\textbf {\bibinfo {volume} {60}},\ \bibinfo {pages} {3107}
  (\bibinfo {year} {1999})}\BibitemShut {NoStop}%
\bibitem [{\citenamefont {Rahman}(1964)}]{ngauss:rahman}%
  \BibitemOpen
  \bibfield  {author} {\bibinfo {author} {\bibfnamefont {A.}~\bibnamefont
  {Rahman}},\ }\href {\doibase 10.1103/PhysRev.136.A405} {\bibfield  {journal}
  {\bibinfo  {journal} {Phys. Rev.}\ }\textbf {\bibinfo {volume} {136}},\
  \bibinfo {pages} {A405} (\bibinfo {year} {1964})}\BibitemShut {NoStop}%
\bibitem [{\citenamefont {Zwanzig}(2001)}]{Zwanzig}%
  \BibitemOpen
  \bibfield  {author} {\bibinfo {author} {\bibfnamefont {R.}~\bibnamefont
  {Zwanzig}},\ }\href@noop {} {\emph {\bibinfo {title} {Nonequilibrium
  Statistical Mechanics}}}\ (\bibinfo  {publisher} {Oxford University Press},\
  \bibinfo {year} {2001})\BibitemShut {NoStop}%
\bibitem [{\citenamefont {Reinhardt}\ \emph {et~al.}(2013)\citenamefont
  {Reinhardt}, \citenamefont {Weysser},\ and\ \citenamefont
  {Brader}}]{0295-5075-102-2-28011}%
  \BibitemOpen
  \bibfield  {author} {\bibinfo {author} {\bibfnamefont {J.}~\bibnamefont
  {Reinhardt}}, \bibinfo {author} {\bibfnamefont {F.}~\bibnamefont {Weysser}},
  \ and\ \bibinfo {author} {\bibfnamefont {J.~M.}\ \bibnamefont {Brader}},\
  }\href {http://stacks.iop.org/0295-5075/102/i=2/a=28011} {\bibfield
  {journal} {\bibinfo  {journal} {EPL}\ }\textbf {\bibinfo {volume} {102}},\
  \bibinfo {pages} {28011} (\bibinfo {year} {2013})}\BibitemShut {NoStop}%
\bibitem [{\citenamefont {D.~Stauffer}(1994)}]{REF12}%
  \BibitemOpen
  \bibfield  {author} {\bibinfo {author} {\bibfnamefont {A.~A.}\ \bibnamefont
  {D.~Stauffer}},\ }\href@noop {} {\emph {\bibinfo {title} {Introduction To
  Percolation Theory}}}\ (\bibinfo  {publisher} {Taylor and Francis,
  Philadelphia},\ \bibinfo {year} {1994})\BibitemShut {NoStop}%
\bibitem [{\citenamefont {Vicsek}(1989)}]{vicsek}%
  \BibitemOpen
  \bibfield  {author} {\bibinfo {author} {\bibfnamefont {T.}~\bibnamefont
  {Vicsek}},\ }\href@noop {} {\emph {\bibinfo {title} {Fractal Growth
  Phenomena}}}\ (\bibinfo  {publisher} {World Scientific Publishing Co. Ptr.
  Ltd., Singapore},\ \bibinfo {year} {1989})\BibitemShut {NoStop}%
\bibitem [{\citenamefont {Poon}\ \emph {et~al.}(1995)\citenamefont {Poon},
  \citenamefont {Pirie},\ and\ \citenamefont {Pusey}}]{Poon:1995ts}%
  \BibitemOpen
  \bibfield  {author} {\bibinfo {author} {\bibfnamefont {W.}~\bibnamefont
  {Poon}}, \bibinfo {author} {\bibfnamefont {A.~D.}\ \bibnamefont {Pirie}}, \
  and\ \bibinfo {author} {\bibfnamefont {P.~N.}\ \bibnamefont {Pusey}},\
  }\href@noop {} {\bibfield  {journal} {\bibinfo  {journal} {Faraday
  Discussions}\ }\textbf {\bibinfo {volume} {101}},\ \bibinfo {pages} {65}
  (\bibinfo {year} {1995})}\BibitemShut {NoStop}%
\bibitem [{\citenamefont {Fortini}(2012)}]{fortini:pickering}%
  \BibitemOpen
  \bibfield  {author} {\bibinfo {author} {\bibfnamefont {A.}~\bibnamefont
  {Fortini}},\ }\href@noop {} {\bibfield  {journal} {\bibinfo  {journal} {Phys.
  Rev. E}\ }\textbf {\bibinfo {volume} {85}},\ \bibinfo {pages} {040401(R)}
  (\bibinfo {year} {2012})}\BibitemShut {NoStop}%
\bibitem [{Note1()}]{Note1}%
  \BibitemOpen
  \bibinfo {note} {In the limit of low densities and large interaction
  strengths the system will reach a fractal dimension $d_f \simeq 1.7$, typical
  of systems formed from diffusion-limited cluster aggregation
  (DLCA)}\BibitemShut {NoStop}%
\bibitem [{\citenamefont {Keddie}\ and\ \citenamefont
  {Routh}(2010)}]{Keddie:2010ta}%
  \BibitemOpen
  \bibfield  {author} {\bibinfo {author} {\bibfnamefont {J.~L.}\ \bibnamefont
  {Keddie}}\ and\ \bibinfo {author} {\bibfnamefont {A.~F.}\ \bibnamefont
  {Routh}},\ }\href@noop {} {\emph {\bibinfo {title} {{Fundamentals of Latex
  Film Formation}}}},\ Processes and Properties\ (\bibinfo  {publisher}
  {Springer},\ \bibinfo {year} {2010})\BibitemShut {NoStop}%
\end{thebibliography}

%

\end{document}